\documentclass[12pt]{iopart}
\usepackage{amssymb,amsfonts}
\usepackage{dcolumn}
\usepackage{bm}
\usepackage{graphicx}
\usepackage{color}
\providecommand{\G}{\mathcal{G}}
\providecommand{\ef}{\varepsilon_F^\ell}
\providecommand{\TF}{T_F^\ell}

\begin{document}

\title{Electron-electron interaction effects in quantum point contacts}
\author{A.M. Lunde,$^{1,2}$ A. De Martino,$^{3,4}$
A. Schulz,$^3$ R. Egger,$^3$\\ and K. Flensberg$^2$}

\address{ $^1$ D\'{e}partment de Physique Th\'{e}orique, Universit\'{e} de Gen\`{e}ve,
CH-1211 Gen\`{e}ve 4, Switzerland\\ $^2$~Nano-Science Center,
Niels Bohr Institute, University of Copenhagen, DK-2100
Copenhagen, Denmark \\$^3$~Institut f\"ur Theoretische Physik,
Heinrich-Heine-Universit\"at,
D-40225  D\"usseldorf, Germany\\
$^4$~Institut f\"ur Theoretische Physik, Universit\"at zu K\"oln,
D-50937  K\"oln, Germany }

\date{\today}

\begin{abstract}
We consider electron-electron interaction effects in quantum
point contacts on the first quantization plateau, taking into
account all scattering processes. We compute the
low-temperature linear and nonlinear conductance, shot noise,
and thermopower, by perturbation theory and a self-consistent
nonperturbative method. On the conductance plateau, the
low-temperature corrections are solely due to
momentum-nonconserving processes that change the relative
number of left- and right-moving electrons. This leads to a
suppression of the conductance for increasing temperature or
voltage. The size of the suppression is estimated for a
realistic saddle-point potential, and is largest in the
beginning of the conductance plateau. For large magnetic field,
interaction effects are strongly suppressed by the Pauli
principle, and hence the first spin-split conductance plateau
has a much weaker interaction correction.  For the
nonperturbative calculations, we use a self-consistent
nonequilibrium Green's function approach, which suggests that
the conductance saturates at elevated temperatures.  These
results are consistent with many experimental observations
related to the so-called 0.7 anomaly.
\end{abstract}

\maketitle

\section{Introduction}
\label{sec:intro}

Conductance quantization in a quantum point contact (QPC),
first observed in 1988\cite{wharam}, constitutes a classic
textbook effect of mesoscopic physics.  On top of the integer
conductance plateaus $G=nG_0$ (where $G_0=2e^2/h$) observed as
a function of gate voltage $V_g$, many experiments have found a
temperature-dependent suppression of the conductance appearing
in the first half of the conductance plateau.  This
shoulder-like feature is seen at elevated temperature $T$ (or
finite voltage $V$) near the first quantized
plateau\cite{thomas,kristensen,cronenwett}, accompanied by a
shot noise reduction\cite{roche}. It appears approximately
around 0.7 $G_0$, and has therefore been named the "0.7
anomaly"\cite{thomas,kristensen,cronenwett,appleyard,thermo2,roche,sfigakis}.
Despite the conceptual simplicity of a QPC and the fact that
the 0.7 anomaly has been observed in a variety of material
systems by different groups over more than a decade, still no
generally accepted microscopic theory exists, apart from an
overall consensus that one is dealing with some spin-related
effect.

While phenomenological models\cite{bruus}, assuming the
existence of a density-dependent spin gap, can provide rather
good fits to experimental data, the presumed static spin
polarization due to interactions within the {\sl local}\ QPC
region is not expected in the presence of unpolarized {\sl
bulk}\ reservoirs. Along this line of thinking, it was recently
conjectured that spin symmetry-broken mean-field or density
functional theory calculations are unable to recover the
correct $T$ dependence of the
conductance\cite{richter,zou,bulka}. A number of microscopic
theories assume the existence of a quasi-bound state in the QPC
region, leading to a Kondo-type scenario, as encountered in
transport through interacting quantum
dots\cite{meir,cornagliabalseiro}. Such a quasi-bound state was
indeed found in spin density functional theory (SDFT)
calculations\cite{meir}, and models based on this picture
appear to reproduce several essential observations related to
the 0.7 anomaly. However, different SDFT works also reached
different conclusions\cite{zou,berggren}. Further proposals
involve phonon effects\cite{seeligmatveev}. Several
publications have suggested that electron-electron (e-e)
interactions alone may already result in a reduced conductance
in a QPC at elevated temperatures, without the need for
additional assumptions of spin polarization or a localized
state\cite{matveev,meidanoreg,schmeltzer,syljuaasen,sushkov,matv2,preprint}.

\subsection{Main ideas and results}

\begin{figure}
\hspace{2.5cm}\includegraphics[width=0.8\textwidth]{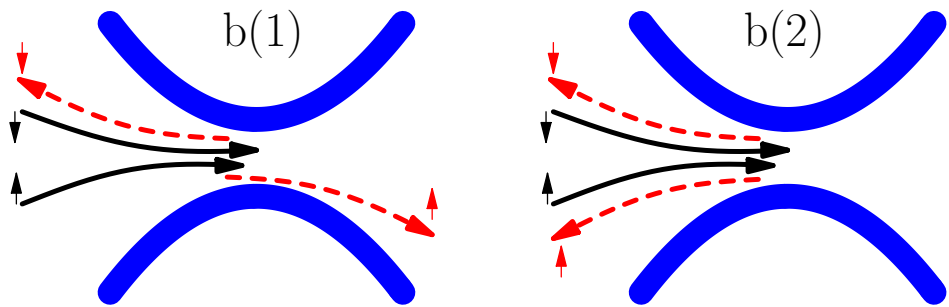}
\caption{ \label{fig:back} Illustration of the
two-electron momentum-nonconserving scattering processes that
give rise to a correction to the transport properties at the
beginning of the first plateau. The full (black) lines
represent incoming electrons, while the dashed (red) lines are
the outgoing electrons. The thick (blue) lines define the edge
of the QPC. Only scattering between different spins is present
to leading order in $T/\TF$ due to the Pauli principle.}
\end{figure}
\label{sec:main}

Motivated by this body of experimental and theoretical work, we
reconsider the role of e-e interactions for electronic
transport properties in  QPCs, starting from the assumption
that in the low-temperature limit, a QPC is well described by a
single-particle saddle-point potential. We then include e-e
scattering, and, in particular, all {\sl
momentum-nonconserving} processes, where the number of left-
and right-moving electrons does not have to be conserved in the
scattering process, see figure \ref{fig:back}). These processes
are not allowed in infinitely long translationally invariant
(single-mode) quantum wires. However, the lack of translational
invariance for a QPC connected to reservoirs permits such
processes here. In other words, momentum is not a good quantum
number for a QPC, and therefore interactions processes can
violate momentum conservation. Momentum-nonconserving processes
are most relevant in the low-density regime, where the Fermi
wavelength is comparable to the QPC's length, which is set by
the curvature of the saddle-point potential and/or the distance
to the gate electrodes\cite{newfoot}. Indeed, our quantitative
analysis of the matrix elements for these processes (see below)
shows that the effect of momentum-nonconserving scattering can
be substantial, and implies that the conductance is
significantly reduced at elevated temperatures, where the phase
space for inelastic scattering is increased. We find that the
breaking of translational invariance, and hence the
backscattering rate, is most dramatic near the onset of the
plateau, and then gradually decreases for larger electron
density in the QPC.

We start from the assumption that the low-temperature limit of
a QPC at the first quantized plateau is well described by a
Fermi-liquid picture with propagating single-particle
states\cite{glazman88}. Throughout the paper, we consider only
a single transverse channel being transmitted. Without
interactions and at low temperatures, $T\ll \TF$ (where
$\ef=k_B\TF$ is the \textit{local} Fermi energy, see
(\ref{localEF}) below), the conductance is given by the
standard Landauer-B$\ddot{\textrm{u}}$ttiker\cite{buttiker86}
formula, $I^{(0)}=G_0\mathcal{T}_0(E_F) V$, where $V$ is the
voltage difference across the contact and $G_0=2e^2/h$. In this
paper, we are interested in the properties at the plateau, i.e.
when the zero-temperature transmission probability is close to
one, ${\cal T}_0 \simeq 1$.  At zero temperature, electrons do
not experience inelastic scattering from either phonons or
other electrons. However, as temperature increases, phase space
also increases for such scattering events. The effect of phonon
scattering has previously been studied by Seelig and
Matveev\cite{seeligmatveev}. Here, we address the question of
inelastic e-e scattering, which can cause electron
backscattering. For example, two incoming electrons from the
high-bias side can interact in the contact and scatter, so that
one electron is backscattered while the other is transmitted.
These processes are later denoted as $b(1)$ and shown in figure
\ref{fig:back}. Also shown is a backscattering process for two
electrons\cite{meidanoreg}, which we denote as $b(2)$. Both
processes will lead to a current reduction.

\begin{figure}
\hspace{4cm}\includegraphics[width=0.4\textwidth]{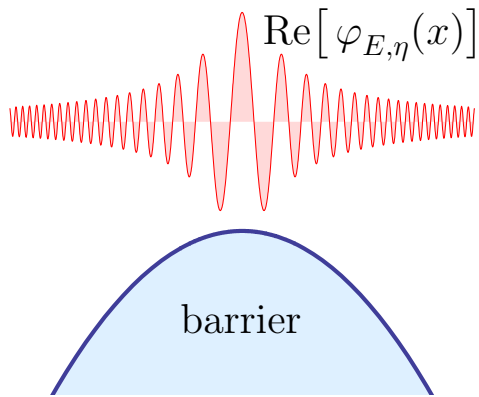}
\caption{\label{fig:wkb} Illustration of the effective 1D
potential of the QPC and the transmitting wavefunction
$\varphi_{E,\eta}$. The wavefunction shows an enhanced weight
in the contact region, which loosely speaking reflects the fact
that electrons spend more time there. This so-called
semiclassical ``slowing down" is the main reason for the finite
values of the backscattering processes shown in
figure \ref{fig:back}, cf. also~\cite{sushkov}, and
will be discussed in section \ref{sec2:B}.}
\end{figure}

The matrix elements for the momentum-nonconserving processes
can, to leading order in the single-particle reflection
$\mathcal{R}_0=1-\mathcal{T}_0$, be calculated using the fully
transmitting wavefunctions. To  good approximation, these can
be described as  WKB states in an effective 1D potential
$V_0(x)$, which is a combination of the potential barrier and
the confinement barrier\cite{glazman88}, see section
\ref{sec:model} and figure \ref{fig:wkb}.  The WKB states at
energy $E$ have the form
\begin{equation}\label{wkbmain}
 \varphi_{E,\eta}(x)= \sqrt{\frac{m}{2\pi\hbar p(x)}}\exp\left(i\eta\int_0^x dx'\,p(x')/\hbar\right),
\end{equation}
where the local momentum is $p(x)=\sqrt{2m[E-V_0(x)]}$, $m$
denotes the effective mass, and $\eta=\pm$ is the propagation
direction.

For a contact with $\mathcal{T}_0\simeq 1$, we may estimate the
leading-order e-e interaction correction to the current by
using Fermi's golden rule. To that end, we compute the rates
for the two scattering events shown in figure \ref{fig:back}
and add them up, with a weight factor keeping track of the
respective contribution to the current. This gives the e-e
interaction correction to the current in the form ($e>0$)
\begin{eqnarray} \label{fgr}
    I^{(2)}&=&\frac{\pi e}{\hbar}\sum_{12,1'2'}|V(12,1'2')|^2
\frac{\eta_{1'}+\eta_{2'}-\eta_1-\eta_2}{4} \\ \nonumber &\times&
    n_1n_2(1-n_{1'})(1-n_{2'})\delta(E_1+E_2-E_{1'}-E_{2'}),
\end{eqnarray}
where we use the short-hand notation $1=\{ E_1,\eta_1,\sigma_1
\}$, with $\sigma$ being the spin index, and sums run over the
quantum numbers of the scattering states,
\begin{equation}\label{sumdef}
\sum_1=\int dE_1\sum_{\sigma_1=\pm}\sum_{\eta_1=\pm}.
\end{equation}
In~(\ref{fgr}), the occupation factors $n_1$ are given by
Fermi-Dirac distribution functions defined by the respective
reservoirs,
\begin{equation}\label{f0def}
n_1= f^0_{L(R)}(E_1)=\frac{1}{e^{(E_1 -\mu_{L(R)})/k_BT_{L(R)}}+1},
\end{equation}
where one should use $\mu_L$ for right-movers $(\eta_1=+1)$ and
$\mu_R$ for left-movers ($\eta_1=-1$), and we have allowed for
different temperatures in the two leads. Furthermore, in
(\ref{fgr}), the factor $(\eta_{1'}+\eta_{2'}-\eta_1-\eta_2)/4$
accounts for the change of the relative number of right- and
left-moving particles from initial to final states. Equation
(\ref{fgr}) is also found from rigorous perturbation theory in
the Keldysh formalism, see section \ref{sec:perturbation},
where the matrix elements $V(12,1'2')$ are specified in
(\ref{VW}). As expected, they contain both a direct and an
exchange term, which is important for the behavior in a large
spin-splitting magnetic field. The magnitude of the relevant
coupling strengths is carefully estimated in section
\ref{sec2:B}.

Based on (\ref{fgr}), for ${\cal T}_0=1$, we obtain the linear
conductance at low temperatures in the form
\begin{equation}\label{cond}
    G/G_0 = 1+G^{(2)}/G_0 = 1-A_b (\pi T/\TF)^2,
\end{equation}
where a realistic estimate for a typical GaAs QPC indicates
that $A_b\approx 1$ at the beginning of the plateau. The result
(\ref{cond}) has also been reported in \cite{sushkov}, where
the QPC was modeled using a kinetic equation. Their prefactor
$A_b$ is proportional to an unknown ``relaxation time in the
leads'', a quantity that does not appear in our theory. Instead
$A_b$ is directly connected to the inelastic e-e interaction
processes. We find that $A_b$ rapidly decreases when moving
along the plateau, see also figure
\ref{fig:interaction-strengths}. The dimensionless coefficient
$A_b$ includes the effects of both the $b(1)$ and the $b(2)$
backscattering processes.  It is also interesting to consider
the interaction correction at spin-split plateaus, i.e. for
large magnetic fields. Here the energy dependence of the matrix
element $V(12,1'2')$ becomes important for equal-spin
scattering, since \emph{exactly at the Fermi energy} the direct
and exchange terms cancel for equal spins only (see section
\ref{sec:largeBfield}). Hence, e-e scattering corrections in
the spin-polarized case are much smaller. In fact, the leading
contribution turns out to be of order $(T/\TF)^4$, and the
prefactor is smaller than $A_b$. Physically, this can be
understood in terms of the Pauli principle. This qualitatively
agrees with the experimental observation that almost no
conductance suppression at the $e^2/h$ plateau occurs.

Below we also provide results for the interaction corrections
to other experimentally relevant quantities, such as shot
noise, the thermopower, and the nonlinear conductance.

The effect of e-e backscattering is thus most important at the
beginning of the $2e^2/h$ plateau. On the other hand, for
elevated temperatures, it leads to a breakdown of perturbation
theory. Therefore, a crossover to a different type of behavior
must appear. From equation~(\ref{cond}), the temperature scale
for this crossover is expected to be
\begin{equation}\label{Tstar}
    T^*\approx \frac{\TF}{\pi\sqrt{A_b}}.
\end{equation}
Contrary to the usual situation encountered in mesoscopic
physics, the nontrivial question to be answered thus concerns
the high-temperature limit (but still $T\ll \TF$). To
investigate this question, we have studied a local interaction
model by means of a nonequilibrium formalism, employing a
self-consistent Born approximation (SCBA) for the self-energy.
In this case, our numerical results suggest that the linear
conductance approaches a saturation value $g_s=G/G_0$ of order
$g_s\approx 1/2$ at high temperatures. The saturation is
physically due to the fact that e-e scattering processes within
the QPC region fully equilibrate outgoing electrons, which also
suggests that $g_s$ is non-universal and thus depends on the
detailed form of the various e-e couplings.  We mention in
passing that a high-$T$ saturation of the conductance has also
been reported for long 1D wires\cite{matveev,footnote:new},

\subsection{Structure of the paper}

The structure of this article is as follows. In section
\ref{sec:model}, we define the model of the QPC, and provide
estimates for the  parameters involved. In section
\ref{sec:perturbation}, this model is treated by lowest-order
perturbation theory, and the interaction corrections to linear
and nonlinear conductance, shot noise, and thermopower are
computed.  A simplified version of the QPC model with a local
e-e interaction potential is then considered within a
nonequilibrium formalism in section \ref{sec:num}, leading to a
self-consistent numerical approach. This allows us to go beyond
lowest-order perturbation theory, albeit in an approximate
fashion. We briefly conclude in section \ref{sec:discussion}.
Details of the calculations can be found in two appendices. In
intermediate steps, we sometimes set $\hbar=1$.

\section{Model and estimates}
\label{sec:model}

\subsection{Model}

We consider a two-dimensional electron gas (2DEG) with a
single-particle potential $U({\bf x})$ forming the QPC, where
${\bf x}=(x,y)$. Close to the middle of the constriction, ${\bf
x}=(0,0)$, the potential is assumed to be described by a
saddle-point potential\cite{buttiker,glazman88-adiabatisk-QPC}
\begin{equation}\label{Ux}
U(\mathbf{x}) =  U_0 - \frac12 m\omega_x^2 x^2 + \frac12 m\omega_y^2 y^2 ,
\end{equation}
where $x$ ($y$) is along (perpendicular to) the transport
direction of the QPC. The QPC is thus characterized by the
frequencies $\omega_x$ and $\omega_y$, or, equivalently, by the
length scales $\ell_y= \sqrt{\hbar/m\omega_y}$ and $\ell_x=
\sqrt{\hbar/m\omega_x}$. Transforming the 2D Schr\"{o}dinger
equation into 1D, the $y$-direction simply gives transversal 1D
subbands (modes) labeled by $n=0,1,2\ldots$. Therefore the
effective 1D Schr\"{o}dinger equation for mode $n$ has a
potential given by
\begin{equation} \label{eq:effective-1D-saddle-point-potential}
V_n(x)=\hbar\omega_y(n+1/2)+U_0-\frac{1}{2}m\omega_x^2 x^2,
\end{equation}
and the well-known transmission probability through the $n^{\textrm{th}}$ mode is\cite{buttiker}
\begin{equation}
\mathcal{T}_n(E)=\frac{1}{1+\exp[-2\pi(E-\hbar\omega_y(n+\frac{1}{2})-U_0)/\hbar\omega_x]}.
\label{eq:transmission}
\end{equation}
This transmission coefficient leads to the low-temperature conductance quantization in a QPC.

In this paper, we focus on the situation when the Fermi energy
is tuned such that only the lowest transversal state $n=0$
propagates through the constriction, and does so with
transmission close to unity, $\mathcal{T}_0(E_F)\simeq 1$.
Therefore only the first (spin-degenerate) mode $n=0$ is
included in the 1D Hamiltonian $H=H_0+H_I$. The noninteracting
part is
\begin{equation}\label{eq:H0}
H_0 = \sum_{\sigma} \int dx\, \psi_\sigma^\dagger (x)
\left( -\frac{\hbar^2}{2m}\frac{d^2}{dx^2} + V_0(x) \right)
\psi^{}_\sigma(x),
\end{equation}
and the Coulomb interaction gives
\begin{equation}\label{hi}
H_I = \frac12 \sum_{\sigma\sigma'} \int dx dx'
\psi_\sigma^\dagger(x) \psi_{\sigma'}^\dagger(x')
W(x,x') \psi_{\sigma'}^{} (x') \psi_\sigma^{} (x),
\end{equation}
where $\psi_\sigma^\dagger(x)$ ($\psi_\sigma^{}(x)$) is the
creation (annihilation) operator for the $n=0$ mode. The
effective (unscreened) 1D interaction $W(x,x')$ can be found by
integration over the transverse eigenstates in the lowest mode,
$\chi_{n=0}(y)=\exp({-y^2/2\ell_y^2})/(\pi^{1/4}
\sqrt{\ell_y})$. This gives
\begin{equation}\label{eq:bessel-int}
W(x,x') = \frac{e^2}{4\pi\epsilon_0\kappa \sqrt{2\pi} \ell_y}\
 M\left(\frac{(x-x')^2}{4\ell_y^2}\right),
\end{equation}
where $\kappa$ denotes the relative dielectric constant,
$M(x)=e^x K_0(x)$, and $K_0$ is the (zeroth-order) modified
Bessel function of the second kind. This model of the
interaction $W(x,x')$ depends only on the difference $x-x'$,
and therefore appears not to break translational invariance.
However, when including the scattering states, the effective
interaction {\sl will}\ break translational invariance, see the
discussion in section \ref{sec2:B} below.

For energies above the barrier top, the scattering states can
be approximated by the WKB eigenstates (\ref{wkbmain}), and we
can thus expand the 1D fermion operators as
\begin{equation}\label{psiscat}
    \psi_\sigma(x)=\sum_{\eta=\pm1}\int dE\, \varphi_{E,\eta}(x)\,c_{E,\eta,\sigma}^{{}} ,
\end{equation}
where the energy integral is from the top of the
single-particle potential to infinity, and
$c_{E,\eta,\sigma}^{{}}$ is a scattering-state annihilation
operator with anticommutation relations
$\{c_{E,\eta,\sigma}^{{}},c_{E',\eta',\sigma'}^{\dagger}\}
=\delta_{\eta,\eta'}\delta_{\sigma,\sigma'}\delta(E-E')$. The
noninteracting Hamiltonian then effectively becomes
\begin{equation}\label{hoss}
H_0 = \sum_1 E_1 c_1^\dagger c_1^{{}},
\end{equation}
where we again use the short-hand notation $1=\{
E_1,\eta_1,\sigma_1 \}$, see (\ref{sumdef}). Similarly, the
interacting part reads
\begin{equation}\label{hIss}
H_I = \frac{1}{2}\sum_{1,2,1',2'} W_{1'2',12} \, c_{1'}^{\dagger}c_{2'}^\dagger c_{2}^{{}}c_{1}^{{}},
\end{equation}
with the matrix elements
\begin{eqnarray}
    W_{1'2',12}=&\delta_{\sigma_1\sigma_{1'}}\delta_{\sigma_2\sigma_{2'}} \int dx_1 dx_2\ \varphi_{E_{1'},\eta_{1'}}^*(x_1) \varphi_{E_{2'},\eta_{2'}}^*(x_2)
    \nonumber\\
    &\times W(x_1,x_2) \varphi_{E_1,\eta_1}^{{}}(x_1) \varphi_{E_2,\eta_2}^{{}}(x_2).\label{WE}
\end{eqnarray}
These matrix elements are discussed in section \ref{sec2:B}.
For later purposes, it is also useful to introduce the
corresponding matrix elements without the spin factors and
taking the energies at the Fermi surface,
\begin{eqnarray}
    W^{(0)}_{\eta_{1'}\eta_{2'},\eta_1\eta_2}=&
\int dx_1 dx_2\ \varphi_{E_F,\eta_{1'}}^*(x_1)
    \varphi_{E_F,\eta_{2'}}^*(x_2)\nonumber\\
    &\times W(x_1,x_2)
    \varphi_{E_F,\eta_1}^{{}}(x_1)
    \varphi_{E_F,\eta_2}^{{}}(x_2).\label{WE0}
\end{eqnarray}
Throughout the paper, we distinguish between the Fermi energy
$E_F$ in the equilibrium leads and the local Fermi energy $\ef$
in the QPC,
\begin{equation}\label{localEF}
\ef\equiv E_F-U_0-\frac{1}{2}\hbar\omega_y = k_B \TF.
\end{equation}
The single-particle thermal smearing of the conductance plateau occurs on the
energy scale $\ef$, and therefore we are here only interested in effects at lower energies, i.e. $T\ll \TF$.

\subsection{Hartree-Fock approximation}\label{sec2:A}

In order to determine the saddle-point potential (\ref{Ux}),
one should in principle solve for the potential
self-consistently, including the screening by surrounding gate
electrodes and mean-field interaction effects. For the
electrons in the constriction, the latter amount to the
Hartree-Fock approximation,
\begin{equation}\label{HIHF}
H_{I,HF} =  \sum_{\sigma} \int dx \int dx'
V_{HF,\sigma}(x,x')\psi_\sigma^\dagger(x')
 \psi_\sigma^{} (x),
\end{equation}
where the self-consistent Hartree-Fock potential is
\begin{eqnarray}
V_{HF,\sigma}(x,x')&=&\delta(x-x')\int dx'' \sum_{\sigma'}W(x,x'')n_{\sigma'}(x'')\nonumber\\
&-&W(x,x')\langle\psi_{\sigma}^\dagger(x)\psi_{\sigma}(x')\rangle,\label{VHFself}
\end{eqnarray}
with
$n_\sigma(x)=\langle\psi_{\sigma}^\dagger(x)\psi_{\sigma}(x)\rangle$.
The self-consistent mean-field theory was discussed in
\cite{richter}, showing that the mean-field result for the
$T$-dependence of the conductance is markedly different from
both the experimental observations and the finite-$T$
corrections due to inelastic e-e processes studied in this
paper.

To illustrate this point, let us consider the Hartree
approximation. Then the change of the local density $n(x)$ with
increasing temperature follows by using a Sommerfeld expansion.
Noting that the chemical potential is set by the electrodes and
can be assumed constant in this temperature range, we find to
lowest order in temperature
\begin{equation}\label{nhartree}
    n(x)=n_0(x)\left[1-\frac{\pi^2}{24}\left(\frac{k_BT}{\ef+\frac12 m\omega_x^2x^2}\right)^2\right],
\end{equation}
where $n_0(x)$ is the $T=0$ density. First, we note that the
density decreases with increasing temperature, because the 1D
density of states decreases with increasing energy.  When
(\ref{nhartree}) is inserted into the Hartree potential, i.e.
into the first term in (\ref{VHFself}), the temperature
correction yields an additional contribution to the
saddle-point potential $V_0(x)$. Expanding this  result around
the relevant barrier region $x=0$, the potential correction can
be written as
\begin{equation}
V_{H}(x) = \delta U_0 +  \Delta \frac{m \omega_x^2}{2} x^2 + {\cal O}\left( (x/2\ell_y)^4 \right).
\end{equation}
Thereby the Hartree correction can be captured by a
temperature-dependent renormalization of the barrier height
$U_0\to U_0+\delta U_0(T)$ and  of the curvature $\omega_x \to
\sqrt{1-\Delta(T)}\omega_x$ of the saddle potential in the
transport direction. Explicitly, in terms of Meijer's
$G$-function\cite{gry}, we find
\begin{equation}\label{du}
\delta U_0 = - \overline W \left(\frac{k_B T}{\hbar \omega_x}\right)^2 \left(
\frac{\hbar\omega_x}{\ef}\right)^{3/2} G^{3,2}_{2,3}\left(\frac{\omega_y\epsilon^\ell_F}{\hbar\omega^2_x} \left|
\begin{array}{lll} 1 & 1 & \\ \frac{1}{2} & \frac{1}{2} & \frac{3}{2} \end{array} \right.  \right) ,
\end{equation}
with the energy scale
\begin{equation}\label{wbar}
\overline W = \frac{1}{\sqrt{18\pi}} \frac{e^2}{4\pi \epsilon_0\kappa \ell_x}.
\end{equation}
Moreover, the dimensionless parameter $\Delta$ follows as
\begin{eqnarray} \label{deltadef}
\Delta&=& \frac{2}{15} \frac{\omega_y}{\omega_x}
\frac{\overline W}{\hbar \omega_x} \left(\frac{k_BT}{\hbar\omega_x}\right)^2
\left(\frac{\hbar\omega_x}{\epsilon_F^\ell}\right)^{\frac{5}{2}} \\
\nonumber &\times& \left[
G^{3,2}_{2,3}\left(\frac{\omega_y\epsilon^\ell_F}{\hbar\omega^2_x} \left| \begin{array}{lll} 1 & 1 & \\
\frac{1}{2} & \frac{1}{2} & \frac{7}{2} \end{array} \right. \right)\right.
%\\\nonumber &&
\left.-4G^{3,2}_{2,3}\left(\frac{\omega_y\epsilon^\ell_F}
{\hbar\omega^2_x} \left| \begin{array}{lll} 1 & 0 & \\
\frac{1}{2} & \frac{1}{2} & \frac{5}{2} \end{array} \right.  \right) \right].
\end{eqnarray}
The barrier is therefore lowered, $\delta U_0<0$, while the
frequency $\omega_x$ is renormalized to smaller values, since
(\ref{deltadef}) implies $\Delta(T)>0$.  The quoted expressions
are useful for ${\cal T}_0\simeq 1$, where $\ef
> \hbar\omega_x/2$.  Inspection of
(\ref{eq:transmission}) then shows that the net effect of the
Hartree correction is an {\sl enhancement}\ of the transmission
towards ${\cal T}_0=1$.  As a consequence, a conductance
plateau present already at zero temperature is then essentially
not affected by the Hartree contribution as long as $T\ll \TF$.
A quantitative discussion of the Hartree corrections to the
saddle-point potential can be found in section \ref{sec2:B}
below.

Concerning the Fock part, a simple qualitative approximation is
obtained if the pair potential is assumed to be a contact
interaction, $W(x,x')=W_0 \delta(x-x')$, see also
\cite{richter}.  (This expression requires that the long-ranged
tail of the e-e interaction potential is screened by nearby
electrodes.) In that case, also the Fock contribution is local,
and acts in precisely the same (but opposite) way as the
Hartree potential, thereby partially cancelling the effects of
the latter.  In particular, it leads to the same temperature
dependence of the corrections.  We therefore conclude that the
Hartree-Fock contribution leads to a very weak temperature
dependence of the conductance, which even goes in opposite
direction as compared to the effects of the inelastic e-e
contributions.  This temperature dependence of the mean-field
result for the QPC conductance has been described in detail
before\cite{richter,zou}.

In the following, the weak $T$-dependence of the conductance
under a single-particle picture will be neglected, and we focus
on  the effects of inelastic e-e scattering. Hence the
interaction (\ref{hi}) is replaced by
\begin{equation}\label{replace}
\delta H_I=H_I-H_{I,HF}.
\end{equation}
When performing diagrammatic expansions, the first-order
contribution in $\delta H_I$ vanishes per definition, and the
series starts with the second order in $\delta H_I$, see
sections \ref{sec:perturbation} and \ref{sec:num}.  {}From now
on, the Hartree-Fock part is assumed to be contained in the
single-particle potential $V_0(x)$.

\subsection{WKB estimates for e-e matrix elements} \label{sec2:B}

In order for our subsequent calculations to be meaningful, it
is essential to first show that the relevant e-e backscattering
strengths can be sufficiently large in experimentally relevant
QPC geometries.  In the following, we therefore provide such
estimates and demonstrate that in the beginning of the first
conductance quantization plateau, these couplings are indeed
significant.  Let us then consider an almost perfectly
transmitting QPC and calculate the matrix elements that give
rise to momentum-nonconserving scattering. For a simple
estimate of the relevant e-e backscattering strengths, we now
employ the WKB wavefunctions, see section \ref{sec:main}. For
larger reflection probability the WKB approximation is not
adequate and a more sophisticated approach must be used.

Because of the lower electron density in the QPC region, also
denoted as {\sl semiclassical slowing down}\cite{sushkov},
interactions are strongest near the constriction, see figure
\ref{fig:wkb}. The single-particle eigenstates in the potential
$V_0(x)$ are given by the WKB scattering states
(\ref{wkbmain}), which are a good approximation for
$\mathcal{T}_0(E_F)\simeq 1$. The following estimates can thus
be regarded as an expansion in $\mathcal{R}_0=1-\mathcal{T}_0$,
to leading order in $\mathcal{R}_0<<1$. In the WKB
approximation, the matrix elements (\ref{WE0}) for e-e
scattering at the Fermi surface are given by
\begin{eqnarray}
 W^{(0)}_{\eta_{1'}\eta_{2'},\eta_1\eta_2}&=& \left(\frac{m}{2\pi\hbar}\right)^2 \int dx dx'\, \frac{W(x,x')}{p(x)p(x')}
\label{Wdef}
\nonumber \\ && \times \,
e^{i(\eta_1-\eta_{1'})\int_0^{x} dy\,p(y)/\hbar} \,
    e^{i(\eta_2-\eta_{2'})\int_0^{x'} dy\,p(y)/\hbar}.
\end{eqnarray}
The semiclassical    slowing down is reflected in the local
density of states $\propto 1/p(x)$, which is largest inside the
QPC ($x=0$). Here, $p(x)=\sqrt{2m[E_F-V_0(x)]}$ is the
classical momentum, which we express using the 1D saddle-point
potential $V_0(x)$ in
(\ref{eq:effective-1D-saddle-point-potential}) as
\begin{eqnarray}
p(x)=\frac{a}{\ell_y}\frac{\hbar}{\ell_y}\frac{\omega _{x}}
{\omega _{y}}\sqrt{1+(x/a)^{2}}.
\end{eqnarray}
The Fermi energy $E_F$ enters here via the length scale $a$,
\begin{equation}\label{aell}
\frac{a}{\ell_y}= \frac{\omega _{y}}{\omega _{x}}\sqrt{\frac{2(E_F-U_0-\hbar \omega_{y}/2)}{\hbar \omega_{y}}}.
\end{equation}
This scale can also be related to the transmission
$\mathcal{T}_0(E_F)$ by using (\ref{eq:transmission}), and
hence to the position along the first conductance plateau.
Inverting (\ref{eq:transmission}), we find
\begin{equation}\label{aTrel}
\left(\frac{a}{\ell_y}\right)^2=\frac{1}{\pi}\frac{\omega_y}{\omega_x}\ln\left(\frac{\mathcal{T}_0}{1-\mathcal{T}_0}\right),
\end{equation} valid for $0.5 \leq \mathcal{T}_0<1$. Next,
consider the integral in the phase factor of (\ref{Wdef}),
\begin{eqnarray}
\gamma(\xi) &\equiv\frac{1}{\hbar }\int_{0}^{x}dy~p(y) =-\gamma(-\xi)\\
&=\frac{1}{2\pi}\ln\!\left[\frac{\mathcal{T}_0}{1-\mathcal{T}_0}\right]
\left[\xi \sqrt{\xi ^{2}+1}+\ln \left(\xi +\sqrt{\xi ^{2}+1}\right)\right],\nonumber
\end{eqnarray}
where $\xi=x/a$. This expresses (\ref{Wdef}) in a form suitable
for numerical integration,
\begin{equation}\label{Wf}
  W^{(0)}_{\eta_{1'}\eta_{2'},\eta_1\eta_2}= \frac{1}{(2\pi \hbar\omega_x)^2 } \frac{e^{2}}{4\pi\epsilon_0
\,\kappa\ell_y}f_{\eta_{1'},\eta_{2'},\eta_1,\eta_2}
\end{equation}
where
\begin{eqnarray}
f_{\eta_{1'},\eta_{2'},\eta_1,\eta_2}=&\frac{1}{\sqrt{2\pi}}\int_{-\infty}^\infty
d\xi \int_{-\infty}^\infty d\xi^{\prime} \frac{1}{\sqrt{\xi^{2}+1}} \frac{1}{\sqrt{\xi^{\prime2}+1}} \nonumber\\
\label{fdef} &\times M\!\left[\frac{(\xi -\xi ^{\prime})^{2}}{4}\left(\frac{a}{\ell_y}\right)^{2}\right] \\ \nonumber
&\times \cos[(\eta_1-\eta_{1'})\gamma(\xi)+(\eta_2-\eta_{2'})\gamma(\xi')]
\end{eqnarray}
with $M(x)$ defined after (\ref{eq:bessel-int}). The matrix
elements for the $b(1,2)$ scattering processes in figure
\ref{fig:back} correspond to
\begin{equation}\label{wdef}
W_{b(1)}=W^{(0)}_{++,+-}, \quad W_{b(2)}=W^{(0)}_{++,--},
\end{equation}
while e-e forward-scattering processes are described by $W_{b(0)}=W^{(0)}_{++,++}$.
The couplings $W_{b(i)}$ enter the current through the dimensionless parameters
\begin{eqnarray}
%\nonumber
 A_{b(i)}&=& \frac{16\pi^2}{3}  ( k_B \TF W_{b(i)})^2
\label{Aest}
%\\&=&
=\frac{1}{12\pi^2}\left[ \frac{1}{\hbar\omega_y }
\frac{e^{2}}{4\pi\epsilon_0 \,\kappa\ell_y}
\left(\frac{a}{\ell_y}\right)^2 f_{b(i)} \right]^2.
\end{eqnarray}

\begin{table}
\hspace{2.5cm}\begin{tabular}{|c|c|c|c||c|c|c|} \hline
$\ef$ & $\TF$ & $a/\ell_y$ & $\mathcal{T}_0(E_F)$ & $f_{b(1)}$ & $  W_{b(1)}$ & $A_{b(1)}$ \\
\hline
$\frac{2}{9}\hbar\omega_y$ &  2.3 K & 2.0 &    0.985 & 0.64               & 720 eV$^{-1}$ & 1.1 \\
$\frac{1}{2}\hbar\omega_y$ &  5.2 K & 3.0 &    1.0   & $3.0\times10^{-2}$ &  34 eV$^{-1}$ & $1.0\times10^{-2}$ \\
$\frac{7}{9}\hbar\omega_y$ &  8.1 K & 3.7 &    1.0   & $1.0\times10^{-3}$ & 1.1 eV$^{-1}$ & $1.5\times10^{-5}$\\
\hline
\end{tabular}
\caption{\label{table:Ab1} Estimates for the interaction matrix
elements $  W_{b(1)}$ and $A_{b(1)}$, see (\ref{Aest}), at
three different positions on the first conductance plateau,
i.e.~versus the local Fermi energy $\ef$. Parameters are:
$\hbar\omega_x=0.3$ meV, $\hbar\omega_y=0.9$ meV, $\kappa=10$
and $m=0.067m_e$. See also figure
\ref{fig:interaction-strengths}.}
\end{table}

\begin{figure}
\hspace{3.5cm}\includegraphics[width=0.6\textwidth]{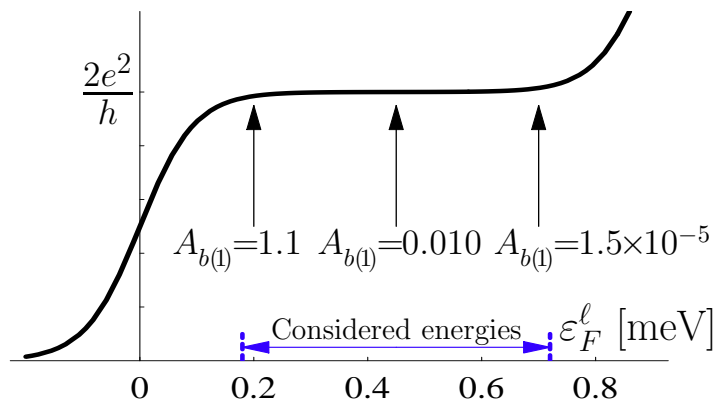}
\caption{\label{fig:interaction-strengths} The
interaction matrix elements for backscattering of a single
electron at various positions on the first quantized
conductance plateau. The solid curve shows the $T=0$
conductance vs $\ef$ for a saddle-point potential with
$\hbar\omega_x=0.3$meV and $\hbar\omega_y=0.9$meV. The decrease
of the conductance in  (\ref{currpert}) is then orders
of magnitude stronger in the beginning of the plateau. The
backscattering of one electron ($A_{b(1)}$) is significantly
stronger than that of two ($A_{b(2)}$). Neglecting $A_{b(2)}$,
the conductance corrections for the three values of $A_{b(1)}$
are: $G^{(2)}/G_0=-2.1\textrm{K}^{-2}\times T^2$,
$G^{(2)}/G_0=-3.6\times10^{-3}\textrm{K}^{-2}\times T^2$ and
$G^{(2)}/G_0=-2.2\times10^{-6}\textrm{K}^{-2}\times T^2$,
respectively. For $T=T^\ast$, we have $G^{(2)}=G_0$, and
perturbation theory fails. This corresponds to $T^\ast=0.7$K,
$T^\ast=17$K, and $T^\ast=670$K for the three stated values of
$A_{b(1)}$. Therefore, it is desirable to go beyond a
perturbative treatment. The quoted $A_{b(1)}$ values are found
numerically by integration of (\ref{fdef}).}
\end{figure}

We now use typical experimental parameters for a GaAs
QPC:\cite{Taboryski1995} $\hbar\omega_x=0.3$ meV,
$\hbar\omega_y=0.9$ meV, $\kappa=10$ and $m=0.067m_e$. We have
numerically integrated $f_{b(1)}$ at three different values of
the local Fermi level $\ef$, see table \ref{table:Ab1}.
Clearly, $W_{b(1)}$ becomes two orders of magnitude smaller
from the beginning of the conductance plateau to the end, while
$A_{b(1)}$ differs even by five orders of magnitude, see figure
\ref{fig:interaction-strengths}.

Let us then briefly discuss the magnitude of the Hartree
corrections mentioned in section \ref{sec2:A}.  Given the QPC
parameters in figure \ref{fig:interaction-strengths}, we obtain
the energy scale $\overline W\approx 0.31$~meV from
(\ref{wbar}). It is then possible to compare the Hartree
conductance correction to the respective inelastic correction
(\ref{cond}). For concreteness, we take $T=0.7~K$,
corresponding to $T^\ast$ at the beginning of the plateau in
figure \ref{fig:interaction-strengths}.  {}From (\ref{du}), at
the positions of the arrows in figure
\ref{fig:interaction-strengths}, we find $\delta U_0\approx
-0.24$~meV, $-0.08$~meV, and $-0.05$~meV (from left to right
along the conductance plateau). Similarly, (\ref{deltadef})
gives $\Delta\approx 0.77, 0.13$, and $0.05$, respectively. The
Hartree corrections to the conductance are then captured by
(\ref{eq:transmission}), and imply an increase of the
conductance.  Since for all three cases in figure
\ref{fig:interaction-strengths}, the transmission at zero
temperature is ${\cal T}_0\geq 0.985$, the Hartree correction
can at most enhance the conductance by a factor $\approx
1.015$, which is a much smaller effect than the inelastic
corrections discussed in this paper. Note that including the
Fock term reduces the found Hartree contribution even further.

Finally,  screening by higher (closed) subbands and by nearby
gate electrodes also leads to an interaction with broken
translational invariance, where the effective interaction range
is set by a combination of the pinch-off distance for closed
subbands and the screening length of the electrodes. This would
have two consequences, namely (i) an overall decrease of the
e-e interaction strength, and thus of the $A_{b(i)}$, and (ii)
an additional source for  broken translational invariance,
which enhances the $A_{b(i)}$. The net result of including
screening by the adjacent gates and 2DEG regions is therefore
not straightforward, and a detailed electrostatic calculation
is needed to gain a reliable understanding.  Since
semiclassical slowing down already results in very significant
backscattering amplitudes, we focus on that in the present
paper.

\section{Perturbation theory}
\label{sec:perturbation}

In this section, we compute the leading interaction corrections
to the transport properties of an almost perfectly transmitting
QPC. As discussed in section \ref{sec:model}, we treat the
interaction Hamiltonian $\delta H_I$ by perturbation theory,
yielding the current as  $I=I^{(0)}+I^{(2)}+\cdots$.  We here
employ the Keldysh formulation to compute the correction
$I^{(2)}$ due to inelastic e-e scattering to second order in
the interaction, and start with the definition of the current
\begin{equation}\label{Idef}
    I=ie\sum_{11'}\langle 1'|\hat{J}(x)|1\rangle \mathcal{G}^<(11',tt),
\end{equation}
where $\mathcal{G}^<(11',tt')=i\langle
c_{1'}^\dagger(t')c_1^{{}}(t)\rangle$ is the nonequilibrium
lesser Green's function (GF), and $\hat
J(x)=\frac{\hbar}{2mi}(\overrightarrow{\partial_{x}}-\overleftarrow{\partial_{x}})$.
For ${\cal T}_0\simeq 1$, we can use the WKB states in
(\ref{wkbmain}) to calculate the current matrix element,
\begin{eqnarray}\label{Jxmatrixelement}
    \langle 1'|\hat{J}(x)|1\rangle&=
   \delta_{\sigma_1\sigma_{1'}} \frac{1}{4\pi\hbar}\frac{\eta_1p_1(x)+\eta_{1'}
    p_{1'}(x)}{\sqrt{p_1(x)p_{1'}(x)}}
    \nonumber\\
    &\times\exp\left(\frac{i}{\hbar}\int_0^x\!\!\!dy[\eta_1p_1(y)-\eta_{1'}p_{1'}(y)]\right).
\end{eqnarray}
Here, we neglect terms proportional to $\partial_x p(x)$ to be
consistent with the WKB approximation.  In order to also allow
for different temperatures in the two leads, we employ the
nonequilibrium formalism in a slightly different way than
usual\cite{Jauho-book,Rammer-smith1986}, namely we include the
different chemical potentials and temperatures in the initial
density matrix $\rho_0(H)$. (The left/right reservoirs inject
electrons distributed by Fermi functions (\ref{f0def}).) The
interaction Hamiltonian $\delta H_I$ is treated as the
perturbation that connects the two reservoirs. To evaluate the
current, we thus write
\begin{eqnarray} \label{eq:operator-average}
I = \textrm{Tr}[\rho(H_0) \hat J(x,t)],
\end{eqnarray}
where $\hat J(x,t)$ is in the Heisenberg picture, $\hat
J(x,t)\equiv e^{iHt/\hbar}\hat J(x)e^{-iHt/\hbar}$. Here
$H=H_0+\delta H_I$, and
\begin{eqnarray} \label{eq:density-matrix}
\rho(H_0)= Z^{-1} e^{-(H_{0,L}-\mu_L
N_{L})/k_BT_L -(H_{0,R}-\mu_R N_{R})/k_B T_R},
\end{eqnarray}
where $N_{L/R}$ and $H_{0,L/R}$  are the number operators and
noninteracting Hamiltonians, respectively,  for electrons
coming from the left/right lead. The noninteracting (retarded,
advanced, lesser, and greater) GFs then follow in the
form\cite{Jauho-book}
\begin{eqnarray}\label{G0s}
  \mathcal{G}_0^{r/a}(1,\omega)&=&\frac{1}{\omega-E_1\pm i0^+}\\ \nonumber
  \mathcal{G}_0^{<}(1,\omega)&=&2\pi i n_1 \delta(\omega-E_1),\\ \nonumber
  \mathcal{G}_0^{>}(1,\omega)&=&-2\pi i[1- n_1]\delta(\omega-E_1),
\end{eqnarray}
where $n_1$ is given in (\ref{f0def}).  The GF's are diagonal
in spin space and identical for both spin directions.

\begin{figure}
\hspace{2.5cm}\includegraphics[width=0.8\textwidth]{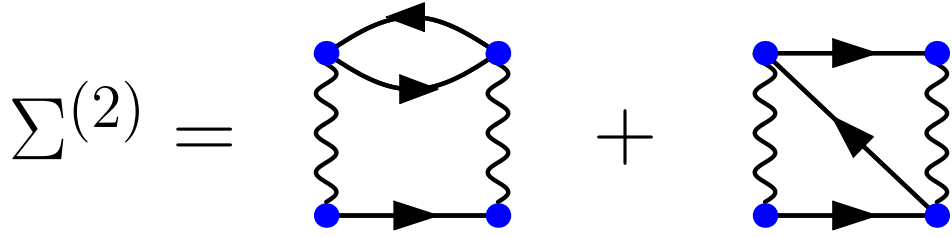}
\caption{ \label{fig:G0} The second-order
diagram included in (\ref{deltaIgen}). The first-order
and all other second-order diagrams belong to the
self-consistent Hartree-Fock sector and are contained in the
saddle-point potential, see section \ref{sec2:A}. To avoid
double-counting, they should therefore not be taken into
account here.}
\end{figure}

We now use a diagrammatic expansion of the interaction operator
in terms of contour-ordered GFs\cite{Jauho-book}. Since the
first-order diagrams vanish per definition, the leading
diagrams are given by the second-order self-energy diagrams
shown in figure \ref{fig:G0}. The second-order correction to
the lesser GF can be found by using standard Langreth
rules\cite{Jauho-book} cf.~(\ref{dyson2}).  After some algebra,
employing (\ref{G0s}), we obtain
%\begin{widetext}
\begin{eqnarray}
I^{(2)}  &=ie\sum_{11^{\prime}}\,\int\frac{d\omega}{2\pi\hbar}
\frac{\langle1^{\prime}|\hat J (x)|1\rangle}{E_{1}-E_{1^{\prime}}-i0^+
}\left\{[\mathcal{G}_{0}^{r}(1,\omega)-\mathcal{G}_{0}^{a}(1^{\prime},\omega)]\Sigma^{(2)<}(11^{\prime},\omega)\right.
\nonumber\\&\left.  -\Sigma^{(2)r}(11^{\prime},\omega)\mathcal{G}_{0}^{<}(1^{\prime}
,\omega)+\mathcal{G}_{0}^{<}(1,\omega)\Sigma^{(2)a}(11^{\prime},\omega)\right\}.
\end{eqnarray}
One checks that this expression conserves current by
differentiating with respect to $x$, and therefore we can set
$x=0$. Now there are two contributions to the current,
corresponding to the real and imaginary parts of
$[E_{1}-E_{1^{\prime}}-i0^+]^{-1}$, where symmetry arguments
show that the real part must vanish.  For the imaginary part,
we have a $\delta$-function enforcing $E_1=E_{1'}$, which then,
see (\ref{Jxmatrixelement}), implies $\eta_1=\eta_{1'}$. This
leaves us with
\[
I^{(2)}  =-\frac{ie}{2\hbar}\sum_{1}\eta_1\left[(1-n_1)\Sigma^{(2)<}(11,E_1)+n_1\Sigma^{(2)>}(11,E_1)\right],
\]
where we use
$\Sigma^{(2)r}-\Sigma^{(2)a}=\Sigma^{(2)>}-\Sigma^{(2)<}$. The
lesser self-energy has two parts, the direct and the exchange
contribution, corresponding to the two terms in figure
\ref{fig:G0}. In the time domain, this gives
\begin{equation}\label{Sigmalesser2nd}
    \Sigma^{(2)<}(11,tt')=-\sum_{234}\mathcal{G}_0^<(2,tt')\mathcal{G}_0^>(3,tt')\mathcal{G}_0^<(4,tt')(W_{13,24}-W_{31,24})W_{24,13}.
\end{equation}
The expression for $\Sigma^{(2)>}$ follows by interchange of
lesser and greater GFs in (\ref{Sigmalesser2nd}). After a
number of manipulations, we reduce this to
\begin{eqnarray}
I^{(2)}=&-\frac{e\pi}{\hbar} \sum_{1,2,1',2'}\left| {W}_{12,1'2'}- {W}_{21,1'2'}\right|^2 \frac{1}{8}(\eta_{1}+\eta_{2}-\eta_{1'}-\eta_{2'})
\nonumber \\
&\times\left[n_{1'}n_{2'}(1-n_{1})(1-n_{2})-(1-n_{1'})(1-n_{2'})n_{1}n_{2}\right]\nonumber \\
&\times\delta(E_{1}+E_{2}-E_{1'}-E_{2'}).
\label{deltaIgen}
\end{eqnarray}
%\end{widetext}
In fact, this result recovers the Fermi golden rule expression
(\ref{fgr}), as may be seen by combining the two terms in
(\ref{deltaIgen}). We can now read off the matrix element
entering (\ref{fgr}),
\begin{equation}\label{VW}
    V(12,1'2')=W_{12,1'2'}- {W}_{21,1'2'},
\end{equation}
which involves the direct and the exchange interaction.

\vspace*{1cm}

\subsection{Conductance at low temperatures and voltages}

The perturbative interaction correction (\ref{deltaIgen}) to
lowest order in temperature $T\ll \TF$ and voltage $eV\ll \ef$
is determined only by the interaction matrix elements
(\ref{WE0}) taken at the Fermi surface. Some algebra then gives
\begin{eqnarray}\label{currpert}
 \frac{I(V,T)}{G_0 V} &=&
\mathcal{T}_0(E_F)-\left(A_{b(1)}+A_{b(2)}\right) \left
(\frac{\pi T}{\TF}\right)^2
\\  \nonumber &-& \left(\frac{A_{b(1)}}{4}+A_{b(2)}\right)
\left(\frac{eV}{\ef}\right)^2 + {\cal O}\left(W^3_{b(i)}\right),
\end{eqnarray}
with the dimensionless coefficients $A_{b(1,2)}$, see
(\ref{Aest}), corresponding to $b(1)$ and $b(2)$ processes for
electrons at the Fermi surface. We stress again that
(\ref{currpert}) holds only to lowest order in
$\mathcal{R}_0=1-\mathcal{T}_0$, i.e. close to unity
transmission at zero temperature. As described in section
\ref{sec:main}, the process $b(2)$ corresponds to the
simultaneous backscattering of two electrons, which has been
discussed on a perturbative level in \cite{meidanoreg}. The
process $b(1)$, where a single electron is backscattered, has
not been studied before. It is here important to note that for
both processes, low-energy scattering is only possible between
opposite spins: When the energy arguments in the matrix element
(\ref{VW}) are taken at the Fermi energy, the direct and
exchange terms cancel each other for equal spins. The parameter
$A_{b(1)}$ is expected to dominate over $A_{b(2)}$, and has
been estimated in section \ref{sec2:B} for a typical GaAs QPC.
We found that $A_{b(1)}$ changes from  $\approx 1$ to $\approx
10^{-5}$ from the beginning of the first quantized conductance
plateau to the end, see figure \ref{fig:interaction-strengths}
and table~\ref{table:Ab1}.

{}From (\ref{currpert}), we observe that interaction processes
that do \emph{not}\ change the number of left- and right-movers
give no contribution to the current correction $I^{(2)}$. The
same conclusion is reached for finite-length quantum wires
using a Boltzmann equation
approach\cite{lundeprl,lundeprb2007}. In particular,
forward-scattering processes do not contribute to $I^{(2)}$,
even though total momentum is not necessarily conserved in
these processes either. Here, the important point is only
whether the number of left- and right-moving electrons is
conserved or not.  However, at higher orders, important at
elevated temperatures, all e-e interaction processes can come
into play, see also Appendix A of \cite{lundeprb2007}.

\subsection{Spin-polarized case}
\label{sec:largeBfield}

Next we consider the effect of interactions for a completely
spin-polarized QPC in a large magnetic field, i.e.~on the
$e^2/h$ plateau.  For the spin-degenerate case, we found above
that to leading order in $T/\TF$, only opposite spins interact
due to the Pauli principle. As a consequence, the conductance
correction is further suppressed for a single spin species, as
we shall show now. In the spin-polarized case, we must take
into account the energy dependence of the interaction matrix
elements $W_{12,1'2'}$, which leads to a stronger power-law
suppression in temperature. The current correction to second
order in the interaction for a single spin species follows from
(\ref{deltaIgen}), and in linear response, we find
%\begin{widetext}
\begin{eqnarray}
I^{(2)}&=& -\frac{eV}{k_BT}\frac{2e\pi}{\hbar} \int dE_1dE_2dE_{1'} dE_{2'}\,
  \left(\big|   V_{b(1)\parallel}\big|^2+ \big|   V_{b(2)\parallel}\big|^2\right)\nonumber \\
&&\times n_1n_2(1-n_{1'})(1-n_{2'}) \
  \delta(E_1+E_2-E_{1'}-E_{2'}) ,
\label{deltaIgen-B-large}
\end{eqnarray}
%\end{widetext}
where the $n_i$ are taken for
$\mu_R=\mu_L=E_F$. The same-spin interaction matrix elements
are
\[
    V_{b(1,2)\parallel} = W_{E_1+\uparrow,E_2+\uparrow,E_{1'}\pm\uparrow,E_{2'}-\uparrow}- W_{E_1\leftrightarrow E_2},
\]
i.e.~a direct term minus an exchange term, where $E_1$ and
$E_2$ are interchanged. Here, the index $E_{1^\prime}+\uparrow$
[$E_{1^\prime}-\uparrow$] refers to the $b(1)$ [$b(2)$]
process.  To lowest order in temperature, the leading term is
found by expanding the $V_{b(1,2)\parallel}$ around the Fermi
energy, since $V_{b(1,2)\parallel}=0$ when all energies are
equal to $E_F$. The low-temperature interaction correction to
the spin-polarized linear conductance $G^{(0)}=e^2/h$ is
obtained as
\begin{equation}\label{intcorrb}
G^{(2)}= -\frac{2e^2}{h}B_{b(1)} \bigg(\frac{\pi T}{\TF}\bigg)^4,
\end{equation}
where the dimensionless parameter
\begin{equation}\label{Bb1def}
B_{b(1)} = \frac{32\pi^2}{15}(k_B \TF)^4
\left|(\partial_{E_1}-\partial_{E_2}){W}_{b(1)\uparrow}|_{E_F}^{{}}\right|^2
\end{equation}
with
${W}_{b(1)\uparrow}=W_{E_1+\uparrow,E_2+\uparrow,E_{1'}+\uparrow,E_{2'}-\uparrow}$
is estimated in \ref{sec:WKB-B}. There, we find
$B_{b(1)}\approx 0.1$ at the beginning of the quantized plateau
(for $\mathcal{T}_0(E_F)=0.985$,  corresponding to the first
arrow in figure \ref{fig:interaction-strengths}). For the
$b(2)$ process, the resulting power-law suppression is even of
higher order, since
\begin{equation}\label{wv2}
\Big|(\partial_{E_1}-\partial_{E_2}){W}_{b(2)\uparrow}|_{E_F}\Big|^2=0,
\end{equation}
where
${W}_{b(2)\uparrow}={W}_{E_1+\uparrow,E_2+\uparrow,E_{1'}-\uparrow,E_{2'}-\uparrow}$.
To get (\ref{wv2}), we use $W_{12,1'2'}=W_{21,2'1'}$, which in
turn follows from  $W(x,x')=W(x',x)$.  In conclusion, an extra
factor $(T/\TF)^2$ suppresses the low-temperature correction at
the spin-polarized $e^2/h$ plateau (i.e. the large-$B$ case) as
compared to the spin-degenerate $2e^2/h$ plateau (i.e.~the
$B=0$ case).

\subsection{Noise}

Next we calculate consequences for another observable, namely
nonequilibrium quantum noise. The zero-frequency shot noise
follows from the (symmetrized) two-point correlation function
of the current operator. Perturbation theory yields for the
backscattering noise power in zero magnetic field
\begin{eqnarray}
S_B(V,T) &=& 2 e  \Bigl[ 2  I_{bs(2)}(V,T) \coth(eV/k_B T)
\nonumber\\ \label{noise} && + I_{bs(1)}(V,T)\coth(eV/2k_B T) \Bigr],
\end{eqnarray}
where $I_{bs(1,2)}$ are the current corrections due to
$W_{b(1,2)}$ quoted in (\ref{currpert}) (defined positive for
$V>0$). This is nothing but the famous Schottky shot noise
relation, encoding the charge of the backscattered particles.
Equation (\ref{noise}) predicts an additional factor of two for
the $b(2)$ contribution, because two electrons are
backscattered in that event\cite{meidanoreg}.

The direct perturbative calculation of the noise to second
order in $\delta H_I$ then yields the full noise power of the
transmitted current in the compact form
\begin{equation}\label{ST}
S_T=S_B+4G_0k_BT-8k_BT \partial_V I_{bs},
\end{equation}
where $I_{bs}=I_{bs(1)}+I_{bs(2)}$.  This perturbative result
has also been obtained in a different context
before\cite{saleur}. In the limit $V\to 0$, one recovers the
expected thermal Johnson-Nyquist noise $4k_BT[G_0+G^{(2)}(T)]$
from (\ref{ST}), with the interaction correction $G^{(2)}$
specified in (\ref{cond}). Note that the last term in
(\ref{ST}) ensures that the correct thermal noise formula is
obtained.

Recent noise measurements on the first quantized plateau were
compared to the corresponding single-particle
picture\cite{roche}, and a reduced noise power was observed at
the conductance anomaly.  Let us now connect (\ref{ST}) to this
discussion.  For that comparison, according to \cite{roche},
one subtracts the thermal noise and defines the excess noise as
\begin{equation}
S_I=S_T-4G(V,T)k_BT .
\end{equation}
 For a noninteracting system, this quantity is given by
\begin{equation}\label{spnoise}
S_I^{SP}=2G_0 \widetilde{\cal R} [eV\coth(eV/2k_BT)-2k_BT]
\end{equation}
to lowest order in the reflection coefficient $\widetilde{\cal
R}$, see reference \cite{roche}. In order to compare our
approach to (\ref{spnoise}), we interpret $I_{bs}/G_0 V$ in a
single-particle picture as an effective reflection probability
$\widetilde{\cal R}$. Of course, this reflection is now mainly
caused by interaction processes, and $\widetilde{\cal R}$ is
generally larger than the true noninteracting value ${\cal
R}_0$. Therefore, the difference between the true excess noise
and its single-particle value is, in this framework, given by
\begin{equation}\label{diff}
   \frac{S_I- S_I^{SP}}{2G_0 eV (T/\TF)^2}=
   -2A_{b(1)}\frac{eV}{k_B T} + A_{b(2)}\, h(eV/k_BT),
\end{equation}
where
\begin{equation}
h(x)=-8x+(\pi^2+x^2)\tanh(x/2).
\end{equation}
{}From this expression, one observes that for $eV<$ 6.507
$k_BT$, regardless of $A_{b(1,2)}$, the measured noise should
always be smaller than predicted by a single-particle analysis.
This situation precisely corresponds to the parameter range of
relevance for the  experimental work  of reference
\cite{roche}, where $eV\lesssim 5 k_BT$.  Our results for the
noise are therefore consistent with the conclusions reached in
reference \cite{roche}.  Unfortunately, however, we cannot
compare to the shot noise line profiles reported in Ref.~7,
because our results are only valid to lowest order in
$\mathcal{R}_0$, and such line profiles would require a
calculation as a function of $\mathcal{R}_0$.

\subsection{Thermopower}

Another experimental observable probing the enhanced phase
space for e-e scattering at $T\neq 0$ is the {\sl thermopower}
$\mathcal{S}(T)$\cite{appleyard,thermodef}, for which
perturbation theory for the spin-degenerate case predicts
\begin{equation}\label{thermo}
\mathcal{S}(T) = \frac{k_B}{e} \frac{2\pi^4}{5}
\left(A_{b(1)}+A_{b(2)}\right) \left(\frac{T}{\TF}\right)^3 .
\end{equation}
Since the noninteracting thermopower is exponentially small
[$\propto \exp(-\TF/T)$] at the conductance plateau, the
interaction correction completely determines the
low-temperature thermopower\cite{lundeprl}. The enhanced
thermopower (as compared to the noninteracting one) is in
qualitative agreement with experiments at the anomalous
plateau\cite{appleyard,thermo2}.

\section{Self-consistent nonperturbative scheme}
\label{sec:num}

As discussed in section \ref{sec:main}, the intermediate
temperature regime
\begin{equation}\label{temprange}
    T^*\lesssim T\ll \TF
\end{equation}
needs to be understood beyond perturbation theory, see the
caption of figure \ref{fig:interaction-strengths} for typical
values of $T^\ast$. Let us first discuss, on a qualitative
level, what happens in limiting cases, where the QPC transport
problem can be solved without invoking approximations. For
instance, when $A_{b(0,1)}=0$ but $A_{b(2)}\ne 0$,
 one can establish correspondence to the problem of a single
impurity in a Luttinger liquid with interaction parameter
$K=2$, which in turn is solvable\cite{gogolin} and leads to the
high-temperature saturation value $g_s=G(T\gg T^*)/G_0=0$ of
the linear conductance.  Physically, this is quite clear: With
strong $b(2)$ interactions, all particles are backscattered for
increasing phase space, and the conductance approaches zero.
Similarly, when $A_{b(1,2)}=0$ but $A_{b(0)}\ne 0$, i.e. when
only forward scattering is present, the conductance is not
affected at all, $g_s={\cal T}_0$ for  ${\cal T}_0\simeq 1$.
(However, for ${\cal T}_0<1$, forward scattering can increase
the conductance, since such processes give an additive
contribution to the usual elastic transmission.)
Unfortunately, the most interesting special case,
$A_{b(0,2)}=0$ but $A_{b(1)}\ne 0$, seems not described by an
exactly solvable model. With only strong $b(1)$ interactions
present, all particles scatter in the QPC, but half of them are
still transmitted. This suggests a reduction of the conductance
by a factor of two, $g_s=1/2$.

In order to make quantitative progress, we now consider a
simplified model. It is given by the noninteracting part
(\ref{eq:H0}) with $V_0(x)=0$. Therefore, we now have the
situation $E_F=\ef\equiv \varepsilon_F=k_B T_F$. Moreover, we
take the interacting part (\ref{hi})  with a local pair
potential, see also \cite{sushkov},
\begin{equation} \label{localpair}
W(x,x') =\widetilde{W} \delta(x) \delta(x').
\end{equation}
The local nature of the pair potential implies that all e-e
interaction processes have the same amplitude, i.e. the
amplitudes for backscattering of one and two electrons and the
forward-scattering amplitudes are all equal,
$A_b=A_{b(0)}=A_{b(1)}=A_{b(2)}$. This is an artefact of the
oversimplified local interaction (\ref{localpair}).
Nevertheless, a study of this problem is useful as it allows
for an explicit calculation of the crossover from the $T^2$
corrections to the high-temperature conductance saturation. The
model cannot be directly translated to the parameters relevant
for a saddle-point QPC, because the dependence of the different
backscattering matrix elements on the Fermi energy is not
captured. Furthermore, because of the locality, the interaction
potential (\ref{localpair}) acts, in accordance with the Pauli
principle, exclusively among opposite spins. Therefore, the
most important effect of a large $B$-field, which consists of
quenching the backscattering amplitudes,
cf.~section~\ref{sec:largeBfield}, is missed.  We then only
consider $B=0$ for the remainder of this section. For
convenience, we express the interaction strength in terms of a
dimensionless parameter,
\begin{equation}
\lambda= m\widetilde{W}/2\hbar^2\pi^{3/2}.
\end{equation}
Writing $\lambda$ in terms of $A_b$, we
find\cite{footnote:lambda} $\lambda=\sqrt{3A_{b}/\pi}$. On the
first part of the plateau, we found $A_{b(1)}\approx 1$, see
section~\ref{sec2:B}, with $T^*/T_F\lesssim 0.3$. This allows
for a meaningful study of the temperature range
(\ref{temprange}) with $\lambda < 1$.

In order to treat the local interaction (\ref{localpair}), we
start from the Dyson equation for the full Keldysh
single-particle GF $\G_\sigma(x,x';\omega)$ for spin $\sigma$,
which is a $2\times 2$ matrix in Keldysh space. Note that the
interaction (\ref{localpair}) does not flip the spin, and
therefore the Dyson equation is
\begin{eqnarray}\label{dyson}
\G_\sigma(x,x'; \omega) &=& \G_{0;\sigma}(x,x';\omega)
%\\ \nonumber &&
+ \G_{0;\sigma}(x,0;\omega) \Sigma_\sigma(\omega) \G_\sigma(0,x';\omega),
\end{eqnarray}
where the self-energy acts only at $x=x'=0.$ Moreover, both
spins enter symmetrically, $\G_\uparrow=\G_\downarrow$, and we
can suppress the spin index. In \ref{appendix:current-formula},
we show that for the interaction (\ref{localpair}) and a
parabolic dispersion, $\varepsilon_k=k^2/2m$, the current can
be expressed in terms of the local spectral function (at
$x=x'=0$) only,
\begin{equation}\label{current}
I=\frac{e}{h} \sum_\sigma \int_0^\infty d\omega
\left[f_R^0(\omega)-f_L^0(\omega)\right]
\frac{A_\sigma(\omega)}{A_0(\omega)},
\end{equation}
where the Fermi functions are defined in (\ref{f0def}), and the
local spectral function is
\begin{equation}
A(\omega)=i[\G^r(\omega)-\G^a(\omega)]=-2 \ {\rm Im} \G^r(\omega)
\end{equation}
with $\G(\omega)\equiv \G(0,0;\omega)$.
The noninteracting spectral function is
\begin{equation}\label{a0}
A_0(\omega)=2\pi d(\omega)= (2m/\omega)^{1/2}\theta(\omega),
\end{equation}
where $d(\omega)$ is the density of states,
and $\theta$ the Heaviside step function.
Note that the real part of the noninteracting
retarded GF $\G^r_{0}(\omega)$
is nonzero at $\omega<0$ for this dispersion relation,
\begin{eqnarray}
\nonumber \G_{0}^r(\omega)
 &=& \int \frac{dk}{2\pi} \frac{1}{\omega-k^2/2m+i0^+}
\\ &=& - \pi d(\omega) \left( \theta(-\omega)+i\theta(\omega) \right).
\label{gfr0}
\end{eqnarray}
In effect, the nonequilibrium current through the interacting
QPC, (\ref{current}), is fully expressed in terms of the local
retarded GF only.

\begin{figure}
\hspace{4cm}\includegraphics[width=0.4\textwidth]{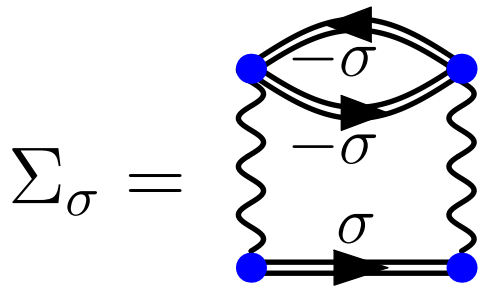}
\caption{ \label{fig:GW} The self-consistent
second-order diagram for the spin direction $\sigma$ included
in our SCBA scheme for the treatment of interactions. Double
lines define full (self-consistent) GFs, as opposed to  thin
lines describing noninteracting GFs, see figure~\ref{fig:G0}.
Note that the exchange term exactly cancels the direct
contribution for same spins, and therefore the second diagram
in figure~\ref{fig:G0} does not appear here. }
\end{figure}

So far, the given relations are exact, but to make progress,
one needs to approximate the self-energy.  As discussed in
section~\ref{sec2:A}, self-consistent Hartree-Fock diagrams are
implicitly included in the single-particle potential $U({\bf
x})$, so we are left with the higher-order diagrams
representing inelastic e-e processes. Here we include the
second-order self-energy in figure~\ref{fig:GW},  made
self-consistent by using the full GF.  Note that the
self-consistency ensures also current
conservation\cite{hershfield}. This diagram amounts to treating
the interactions effectively within the self-consistent Born
approximation (SCBA).  In the perturbative calculations of
section~\ref{sec:perturbation}, the noninteracting GF enters
instead, cf.~figure~\ref{fig:G0}.  In particular, the retarded
component of the self-energy is
\begin{eqnarray}\label{selfen}
\Sigma_\sigma^r(\omega) = \widetilde{W}^2\! \int_0^\infty dt e^{i\omega t}
\left[
\G_{-\sigma}^{<}(-t) \G_{\sigma}^>(t) \G^>_{-\sigma}(t)
%\right. \\ && \left.\quad
 - \G_{-\sigma}^>(-t) \G^<_{\sigma}(t) \G^<_{-\sigma}(t) \right],
\end{eqnarray}
where $\G^{\lessgtr}_\sigma$ denotes the local lesser/greater
GF of the interacting system.  The SCBA approach is similar in
spirit to using a quantum Boltzmann equation approach with a
self-consistent two-body collision integral, see Sec.~6.7.2 in
\cite{lundethesis} for a detailed discussion.

\subsection{Linear transport}
\label{sec4a}

We first discuss the linear conductance, where the spectral
function in (\ref{current}) can be calculated in equilibrium.
Thus, the lesser/greater GFs can be written in terms of the
spectral function,
\begin{equation}\label{Glessgreat}
\G^{<}(\omega) =  i A(\omega) f^0(\omega),\quad
\G^{>}(\omega) = i A(\omega) ( f^0(\omega)-1),
\end{equation}
where $f^0(\omega)=1/[e^{(\omega-\varepsilon_F)/k_B T}+1]$.
Note that when evaluating $\G^{\lessgtr}(t)$ as needed in
(\ref{selfen}),
\begin{equation}\label{ttt}
\G^\lessgtr(t) = \int_{-\infty}^\infty \frac{d\omega}{2\pi}
e^ {-i\omega t} \G^\lessgtr(\omega),
\end{equation}
negative frequencies have to be kept, cf.~(\ref{gfr0}).
Consequently, the self-energy and the interacting spectral
function $A(\omega)$ are generally nonzero at $\omega<0$ due to
the presence of interactions. However, the current only depends
on the spectral functions at $\omega>0$, see (\ref{current}).

To solve the self-consistency problem, we have employed an
iterative procedure.  Starting with the initial guess of the
noninteracting GF, $\G(\omega)=\G_{0}(\omega)$,  see
(\ref{gfr0}) and (\ref{Glessgreat}), we compute
$\Sigma^r(\omega)$ from (\ref{selfen}), which in turn defines a
new retarded GF and a new guess for $A(\omega)$ from the Dyson
equation (\ref{dyson}). In the linear response regime, only the
retarded part of (\ref{dyson}) is needed,
\begin{equation}\label{dysonret}
\G^r(\omega)=\G_{0}^r(\omega)+ \G_{0}^r(\omega)
\Sigma^r(\omega) \G^r(\omega).
\end{equation}
With the solution to (\ref{dysonret}), we then compute a new
estimate for $\Sigma^r(\omega)$, and iterate the procedure
until convergence has been reached. In the numerical
implementation, it is essential to employ fast Fourier
transformation routines to switch between frequency and time
space, cf.~(\ref{ttt}).  This permits the fast evaluation of
the self-energy (\ref{selfen}) in time space. For coupling
strengths $\lambda< 0.8$, this numerical scheme converges to a
unique solution for the spectral functions $A(\omega)$, and can
be implemented in an efficient manner. For larger $\lambda$,
however, several solutions may appear, and the approximation
appears to be ambiguous. We therefore do not show results in
this regime. Given the converged spectral function, we can
compute the linear conductance $G$ from (\ref{current}) by
replacing $f_R^0-f_L^0\to eV[-\partial_\omega f^0(\omega)]$.
Thereby, $G$ is numerically obtained as a function of the
dimensionless parameters $\lambda$ and $T/T_F$.

Let us first briefly discuss the lowest-order result for the
retarded self-energy (\ref{selfen}) at $T=0$.  The result is
obtained by first inserting (\ref{Glessgreat}) with
$A(\omega)\to A_0(\omega)$, see (\ref{a0}), into (\ref{ttt}).
We find
\begin{equation}
\G^<_0(t) = \left( \frac{im}{2\pi^2  t} \right)^{1/2}
 \gamma(1/2,i\varepsilon_F
t),
\end{equation}
with $\G^>_0(t)$ following by overall sign change and
$\gamma\to \Gamma$, with the incomplete gamma functions
$\gamma(\alpha,x)$ and $\Gamma(\alpha,x)$\cite{gry}. For the
chosen dispersion relation, we thereby get an UV-singular
behavior in the self-energy $\Sigma^r(t)$ from (\ref{selfen}),
which manifests itself in the $t\to 0$ behavior
\begin{equation}
\Sigma^r(t\to 0) \simeq - \widetilde{W}^2\theta(t)
\frac{m\sqrt{2m\varepsilon_F}}{\pi^2 t} \left[
1+{\cal O}(t^{1/2}) \right].
\end{equation}
This divergence is also present at finite temperature, but
affects only the real part of the self-energy.  The
perturbative results in section~\ref{sec:perturbation} are thus
insensitive to it. However, when performing a nonperturbative
calculation, a regularization of this UV divergence becomes
necessary, and we chose a bandwidth cutoff of the order
$\omega_c\approx 3\varepsilon_F$ around the Fermi level. It is
numerically convenient to employ a smooth cutoff function (e.g.
a $1/\cosh^2$ filter), but the precise choice for the cutoff
function does not appear to affect our results below.

\begin{figure}
\hspace{2.5cm}\includegraphics[width=0.6\textwidth]{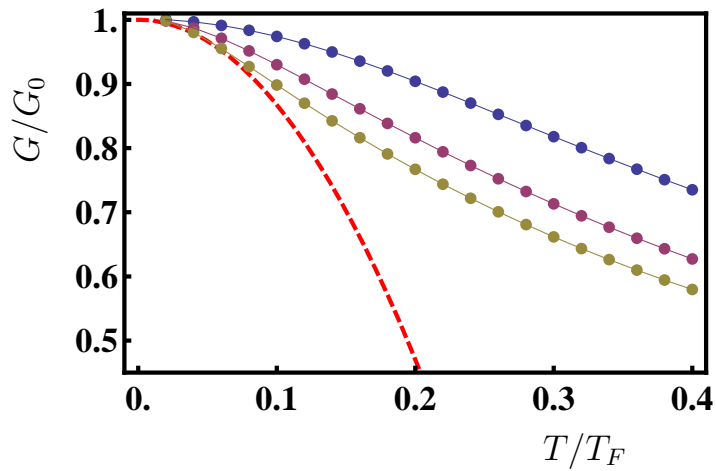}
\caption{ \label{fig:selfratemany} Temperature
dependence of the linear conductance $G$ according to the
self-consistent numerical approach described in
section~\ref{sec4a} for $\lambda=0.3, 0.6,$ and $0.8$ (from top to
bottom curve). The dashed curve gives the perturbative result
for $\lambda=0.8$, dots refer to the numerical data, and solid
curves are guides to the eye only. }
\end{figure}

The numerical results for the linear conductance are shown in
figure \ref{fig:selfratemany}. First of all, we accurately
recover the perturbation theory results at low $T$. At higher
temperatures, a trend towards a conductance saturation could be
conjectured from the numerical data. On a qualitative level,
this saturation value could be expected from the following
simple consideration. In the beginning of
section~\ref{sec:num}, we discussed that including only one of
the backscattering processes of zero, one, or two electrons
could lead to a high-temperature saturation value of $g_s=2,
1,$ or $0$, respectively. Since the point-like interaction
gives the same amplitude to all three processes, it is tempting
to simply take an average of the processes and conjecture a
saturation value of $e^2/h$ for this model.

Different functional forms for $G(T)$ have been used to fit the
experimental data for the 0.7 anomaly, including an activated
$T$-dependence\cite{kristensen} or a  phenomenological
Kondo-type function\cite{cronenwett}. We have tried to fit our
numerical data to both types of functions, slightly generalized
to  allow for different high-$T$ saturation values. The
activated behavior is
\begin{equation}\label{activ}
\frac{G(T)}{G_0} = 1 -(1-g_{s,\lambda}) e^{-T^*_\lambda/T},
\end{equation}
where $g_{s,\lambda}$ characterizes the high-$T$ saturation
value, and $T^*_\lambda$ corresponds to the crossover
temperature scale (\ref{Tstar}).  The Kondo-like equation used
for comparison with experimental data\cite{cronenwett} is a
scaled and shifted version of the universal scaling curve known
from the Kondo problem. (However, this type of modification is
not justified within the Kondo model, where $G\to 0$  at high
$T$.) We can get reasonable fits to both of these functions,
see also \cite{lundethesis}.

As a final remark on the numerical solution of the
self-consistent approach in the linear regime, we mention that
the thermopower (data not shown) exhibits a crossover from the
${\cal S}\propto T^3$ law at low $T$, see (\ref{thermo}), to a
linear-in-$T$ behavior at elevated temperatures.

\subsection{Nonlinear transport}

The iterative solution of the self-consistency problem is also
possible out of equilibrium, and thereby allows to compute the
nonequilibrium current (\ref{current}). The iteration now has
to supplement (\ref{selfen}) for $\Sigma^r$ by the
corresponding equations for the greater/lesser components of
the self-energy,
\begin{equation}
\Sigma^\lessgtr_\sigma(\omega) = \widetilde{W}^2\int_{-\infty}^\infty dt e^{i\omega t} \G_{-\sigma}^\gtrless(-t) \G_{\sigma}^\lessgtr(t)
\G_{-\sigma}^\lessgtr(t) .
\end{equation}
In addition, equation~(\ref{Glessgreat}) holds only in
equilibrium, and has to be generalized. From the Dyson equation
(\ref{dyson}), we obtain the retarded equation (\ref{dysonret})
and (for each $\sigma$) the relation
\begin{eqnarray}\label{dyson2}
&& \G^{\lessgtr}(\omega) = \G_0^{\lessgtr}(\omega) +
 \G_0^r(\omega)\Sigma^r(\omega) \G^{\lessgtr}(\omega)  \\
\nonumber &&+ \G_0^r(\omega)\Sigma^{\lessgtr}(\omega) \G^{a}(\omega) +
 \G_0^{\lessgtr}(\omega)\Sigma^a(\omega) \G^{a}(\omega) ,
\end{eqnarray}
where the advanced components ($\G^a$, $\Sigma^a$) simply
follow from the respective retarded ones by complex
conjugation, since they are defined locally at $x=x'=0$.  For
the iterative procedure, the initial values are now given by
the lesser/greater noninteracting GFs
\begin{eqnarray}\label{gles0}
\G^<_0(\omega) &=& i\pi d(\omega) \theta(\omega) [f^0_R(\omega)+f^0_L(\omega)],
\\ \nonumber \G^>_0(\omega) & =&
i\pi d(\omega) \theta(\omega) [f^0_R(\omega)+f^0_L(\omega)-2].
\end{eqnarray}
The iterative scheme can then be set up in a very similar
manner, and yields a self-consistent solution for the local
spectral function $A(\omega)$, which in turn allows to compute
the current from (\ref{current}). As a check, in the linear
response regime of small bias voltage, we have reproduced the
results of section~\ref{sec4a}.

\begin{figure}
\hspace{2.5cm}\includegraphics[width=0.6\textwidth]{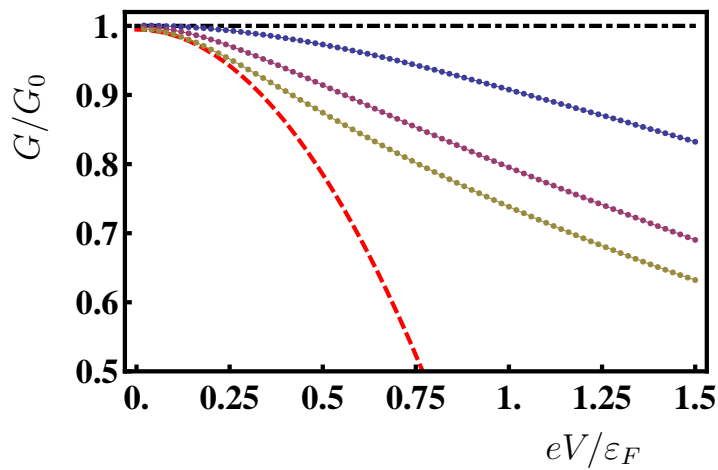}
\caption{ \label{fig:nonlin} Self-consistent
numerical results for the voltage-dependence of the nonlinear
conductance $G=I/V$ for $\lambda=0.3, 0.6, 0.8$ at $T=0.02T_F$
(dotted curves, from top to bottom).  For $\lambda=0.8$, also
the comparison to the perturbative result (dashed curve) is
shown.  For the noninteracting case, $G\simeq G_0$ for
$eV<2\varepsilon_F$ (shown as black dash-dotted curve).}
\end{figure}

Numerical results for the nonlinear conductance $G(V,T)=I/V$ at
a very low temperature are shown in figure \ref{fig:nonlin}. A
clear decrease of the conductance with increasing $V$ is
observed, with a tendency towards saturation of the nonlinear
conductance for large voltages.

\section{Discussion and conclusion}
\label{sec:discussion}

In this paper, we have considered interaction effects in short
QPCs. We have shown that taking into account
momentum-nonconserving processes, we can qualitatively account
for many of the experimentally observed behaviors of the linear
and nonlinear conductance (including their magnetic field
dependencies), thermopower, and shot noise at the 0.7 anomaly.
The gate-voltage dependence can also be qualitatively explained
within the present scheme. The lack of translational invariance
caused by a saddle-point potential is significant for realistic
parameters often used to describe experimental realizations of
QPCs. It is, however, also noteworthy that for longer quantum
wires, where a saddle-point potential is not applicable
anymore, similar backscattering effects can be caused by the
lack of translational invariance at the ends of the quasi-1D
wires.

The self-consistent GF formalism for a simple point-like
interaction model indicates that the conductance may approach a
constant but non-universal value of order $\approx e^2/h$ at
high temperatures.  The precise saturation value is expected to
depend on the detailed e-e backscattering parameters, and hence
on the detailed geometry of the QPC.  The saturation should
appear in the temperature dependence of the linear conductance.
Our numerical data are also consistent with a saturation effect
in the voltage dependence of the nonlinear conductance, but
this issue requires further study.    Physically, the
saturation is caused by a relaxation of the incoming electron
distribution functions due to the interactions within the QPC
region.  It should therefore be possible to develop a
Boltzmann-type kinetic equation approach to describe
(nonequilibrium) inelastic scattering in a QPC, see also
\cite{sushkov} for attempts in this direction. However, such a
development  is not an entirely straightforward procedure and
is outside the scope of this article.

It is also instructive to discuss the point-like interaction
model (\ref{localpair}) in terms of an Anderson impurity model,
see \cite{lundethesis} for details. By spatial discretization
this Hamiltonian maps onto a 1D tight-binding chain with
hopping matrix elements $t$ and on-site interaction $U$ acting
at one site $(x=0$) only. In order to have perfect transmission
at $T=0$ for the Anderson model also, the one-particle on-site
energy must be canceled out so that the local energy is
$\epsilon_0=\varepsilon_F-{\rm Re}\Sigma^r(\varepsilon_F)$. We
thus arrive at an Anderson-type impurity model similar to the
one used in \cite{meir} to describe interactions in a QPC in
the Kondo regime. However, we consider a rather different
parameter regime where $U$ is of the same order as the
hybridization $\Gamma$ and can be parametrically larger than
the bandwidth $D\sim |t|$. Employing (\ref{Tstar}) with
$\ef\approx D$, the interesting temperature range
(\ref{temprange}) translates to $D^2/U < k_B T \ll D$. While
the Kondo model requires the formation of a local moment, this
is not the case for the present approach. It would be
interesting to explore this new parameter regime of the
Anderson model by other non-perturbative schemes developed for
the Anderson model. We believe that our theory offers a
complementary approach to the 0.7 anomaly, which is not based
on the assumption of a (quasi-)bound state in the QPC. In real
samples, however, extrinsic effects (e.g. disorder) may be
responsible for the presence of a bound state, and both
mechanisms may then be relevant in parallel.

To conclude, we hope that our work stimulates additional
theoretical and experimental work on interaction effects in
quantum point contacts. Future theoretical work should analyze
the situation further away from the first conductance plateau,
e.g. when ${\cal T}_0(E_F)$ is not close to
unity\cite{glazman}, or the situation at higher quantization
plateaus. In the latter case, interaction effects due to
semiclassical slowing are in principle also present on the
higher conductance plateaus. However, in this case the other
completely open channels screen the effective interaction,
resulting in a presumably much smaller effect as compared to
the first plateau. While we have repeatedly mentioned the
experimental observations related to the 0.7 anomaly as the
main motivation for our work, it is also clear that a realistic
and quantitative description of experiments needs to consider a
more refined modelling. This is especially important for the
proper description of magnetic field effects away from the
perturbative high magnetic field limit described in section
\ref{sec:largeBfield}. Moreover, on the theoretical side, it
would be highly desirable to go beyond our perturbative
calculation in a more realistic model. For instance, the
gate-voltage dependence of the conductance at elevated
temperatures -- giving rise to the shoulder-like feature
associated with the 0.7 anomaly -- is not accessible to our
theory at the moment, neither in the perturbative regime nor
for the point-like interaction model in section \ref{sec:num},
where the gate-voltage dependence is difficult to incorporate.
In any case, we hope that the model calculations discussed here
can give new insights into the physics behind the observed
anomalies in quantum point contacts.

\ack

We thank P. Brouwer, M. B\"uttiker, W. H\"ausler, J. Paaske,
and E. Sukhorukov for discussions.  A.M.L. appreciates the
support by the Swiss National Science Foundation. We
acknowledge support by the SFB TR 12 of the DFG and by the ESF
network INSTANS.

\appendix

\section{Calculation of $B_{b(1)}$ in the WKB approximation}
\label{sec:WKB-B}

Here we estimate (\ref{Bb1def}) for the prefactor $B_{b(1)}$ of
the $T^4$ interaction correction to the spin-polarized
conductance plateau $e^2/h$, see (\ref{intcorrb}).  The large
$B$-field scattering amplitude involves a derivative of the WKB
states with respect to energy,
\[
\partial_E\varphi_{E,\eta}(x)=\varphi_{E,\eta}(x) \left[-\frac{1}{2}\frac{m}{p^2(x)}+ \frac{i}{\hbar}\eta \int_0^x dy \frac{m}{p(y)}\right].
\]
When inserting this into (\ref{Bb1def}), we obtain
\begin{eqnarray}
( \partial_{E_1}-\partial_{E_2}) {W}_{b(1)\uparrow}|_{E_F}^{{}}
 =&\left( \frac{m}{2\pi\hbar }\right) ^{2}\frac{e^{2}}{4\pi\epsilon_0\sqrt{2\pi}\kappa \ell_y}\int^{\infty}_{-\infty} dx\int^{\infty}_{-\infty} dx^{\prime }~\frac{1}{p(x)p(x^{\prime })}
 \nonumber \\
\nonumber &\times M\left( \frac{(x-x^{\prime })^{2}}{4\ell_y^{2}} \right)\exp \left( \frac{-2i}{\hbar} \int_{0}^{x'}dy\,p(y)\right) \nonumber \\
 &\times \left(\frac{m}{2p^2(x')}-\frac{m}{2p^2(x)}+ \frac{im}{\hbar}\int_{x}^{x'} \frac{dy}{p(y)}\right).
\end{eqnarray}
Using the same parametrization as in section~\ref{sec:model},
this becomes
\[
(\partial_{E_1}-\partial_{E_2}) {W}_{b(1)\uparrow}|_{E_F}= \frac{1}{(2\pi\hbar\omega_x)^2 } \frac{e^{2}}{4\pi\epsilon_0\kappa\ell_y} \frac{1}{\hbar\omega_y}\,g_{b(1)},
\]
where
\begin{eqnarray}\label{g1def}
g_{b(1)}&=& \frac{1}{\sqrt{2\pi}}\int_{-\infty}^\infty
d\xi \int_{-\infty}^\infty d\xi ^{\prime }\frac{1}{\sqrt{\xi^2+1}} \frac{1}{\sqrt{\xi^{\prime2}+1}}M\left[ \frac{(\xi -\xi ^{\prime })^{2}}{4}\left( \frac{a}{\ell_y }
\right)^{2}\right] \nonumber\\ \nonumber
&&\times\left[\frac12\left(\frac{\ell_y}{a}\right)^2
\left(\frac{\omega_y}{\omega_x}\right)^2\left(\frac{1}{1+\xi'^2}-\frac{1}{1+\xi^2}\right) \cos [2\gamma(\xi')]\right.
\nonumber \\
\nonumber &&\left.
\quad+ \frac{\omega_y}{\omega_x}\left[\sinh^{-1}(\xi')-\sinh^{-1}(\xi)\right] \sin[2\gamma(\xi')]\right].
\end{eqnarray}
%\end{widetext}
The value of $B_{b(1)}$ is therefore given by
\[
B_{b(1)}=\frac{1}{120\pi^2} \left(\frac{a}{\ell_y}\right)^8 \left(\frac{\omega_x}{\omega_y}\right)^4 \left[\frac{e^{2}}{4\pi\epsilon_0\kappa\ell_y} \frac{1}{\hbar\omega_y}\ g_{b(1)}\right]^2.
\]

Numerical integration for the same parameters as in
section~\ref{sec2:B} ($\hbar\omega_x=0.3$ meV,
$\hbar\omega_y=0.9$ meV, $\kappa=10$, $m=0.067m_e$, and
$a/\ell_y=2$) then yields $B_{b(1)}\approx 0.1$.

\section{Current for local interaction model}\label{appendix:current-formula}

The purpose of this appendix is to derive (\ref{current}) for
the point-like interaction (\ref{localpair}). The derivation is
a rewriting of the expectation value of the current operator,
see (\ref{eq:operator-average}) and (\ref{eq:density-matrix}),
\[
I(x,t)=\frac{\hbar}{2mi}\sum_{\sigma}
\left[\psi^{\dag}(xt)\big(\partial_x\psi(xt)\big)-\big(\partial_x\psi^{\dag}(xt)\big)\psi(xt)\right],
\]
where $x$ is arbitrary, $\psi^{\dag}(xt)$ is the creation
operator in the Heisenberg picture, and the spin index is
suppressed. In terms of the lesser GF
$\G^{<}(xx',tt')=i\langle\psi^{\dag}(x't')\psi(xt)\rangle$, the
average value is rewritten as
\begin{equation}\label{eq:current-mellem}
\langle I\rangle
=\frac{\hbar}{2m}\sum_{\sigma} \int_{-\infty}^{\infty} \frac{d\omega}{2\pi} \lim_{x'\rightarrow x} \left[\partial_{x'}-\partial_{x}\right]\G_\sigma^{<}(xx',\omega).
\end{equation}
Since the current is independent of $x$, we may take $x=0$ to
evaluate it. To find $\partial_{x}\G^{<}(xx',\omega)$ and
$\partial_{x'}\G^{<}(xx',\omega)$, the Dyson equation is used.
The point-like interaction (\ref{localpair}) allows to simplify
the Dyson equation from an integral equation to an algebraic
one. Using the Langreth rules\cite{Jauho-book}, we obtain for
the lesser GF
\begin{eqnarray}
\G^{<}(xx^{\prime},\omega)=&\G_0^{<}(xx^{\prime},\omega)
+\G_0^{r}(x0,\omega)\Sigma^{r}(00,\omega)\G^{<}(0x^{\prime},\omega)\nonumber\\
&+\G_0^{r}(x0,\omega)\Sigma^{<}(00,\omega)\G^{a}(0x^{\prime},\omega)\nonumber\\
&+\G_0^{<}(x0,\omega)\Sigma^{a}(00,\omega)\G^{a}(0x^{\prime},\omega).\nonumber
\end{eqnarray}
The noninteracting lesser and greater GFs
$\G^{\lessgtr}_0(k,\omega)$ include the Fermi functions of the
leads. In $k$-representation, they are
\begin{eqnarray}
\G^{<}_0(k,\omega)&=+2\pi i\delta(\omega-\varepsilon_k) f_{R/L}^0(\varepsilon_k), \label{eq:G0-lesser}\\
\G^{>}_0(k,\omega)&=-2\pi i\delta(\omega-\varepsilon_k)[1-f_{R/L}^0(\varepsilon_k)], \nonumber
\end{eqnarray}
where $f_{R}^0(\varepsilon_k)$ is used for $k<0$, and
$f_{L}^0(\varepsilon_k)$ for $k>0$. The Dyson equation for
$\G^<$ can also be written with the noninteracting GF $\G_0$ to
the right of the self-energy, e.g.~for $\G^r$, this looks like
$\G^r(xx^{\prime},\omega)=\G_0^r(xx^{\prime},\omega)
+\G^r(x0,\omega)\Sigma^r(00,\omega)\G_0^r(0x^{\prime},\omega)$.
Differentiating the Dyson equation with $\G_0$ on the left
(right) of $\Sigma$ with respect to $x$  ($x'$), we get
derivatives of the noninteracting GFs only,
\begin{eqnarray}
\partial_{x}\G^{<}(00,\omega)=&\partial_{x}\G_0^{<}(00,\omega)\nonumber\\
&+\big[\partial_{x}\G_0^{<}(00,\omega)\big]\Sigma^{a}(00,\omega)\G^{a}(00,\omega),\nonumber\\
\partial_{x'}\G^{<}(00,\omega)=&\partial_{x'}\G_0^{<}(00,\omega)\nonumber\\
&+\G^{r}(00,\omega)\Sigma^{r}(00,\omega)\big[\partial_{x'}\G_0^{<}(00,\omega)\big],\nonumber
\end{eqnarray}
where we use
\[
\lim_{x\rightarrow x'}\partial_{x} \G_0^{r}(xx',\omega) = \int_{-\infty}^{\infty}
\frac{dk}{2\pi} \frac{ik}{\omega-\varepsilon_k +i0^+}  =0,
\]
and likewise for the advanced GF
$\partial_{x}\G_0^{a}(xx',\omega)$, and for differentiation
with respect to $x'$. Inserting this into
\ref{eq:current-mellem}), and using
$\G^{r}(xx',\omega)=[\G^{a}(x'x,\omega)]^{\ast}$,
$\Sigma^{r}(xx',\omega)=[\Sigma^{a}(x'x,\omega)]^{\ast}$, and
$\partial_{x}\G_0(xx',\omega)=-\partial_{x'}\G_0(xx',\omega)$,
we find
\begin{eqnarray}
\langle I\rangle= \frac{\hbar}{m}\sum_{\sigma}
\int_{-\infty}^{\infty}\!\frac{d\omega}{2\pi}
[-\partial_{x}\G_0^{<}(00,\omega)] \,\textrm{Re}\left[1+\G^{r}(00,\omega)\Sigma^{r}(00,\omega)\right].
\end{eqnarray}
Furthermore, from (\ref{eq:G0-lesser}), one obtains
\begin{eqnarray}
\partial_{x}\G_0^{<}(00,\omega)&= \int_{-\infty}^{\infty}\frac{dk}{2\pi}\ \G_0^{<}(k,\omega)\ ik\nonumber\\
&=\theta(\omega) \frac{m}{\hbar^2}\big[f_{R}^{0}(\omega)-f_{L}^{0}(\omega)\big]. \nonumber
\end{eqnarray}
Thereby we can express the current in terms of local GFs only,
$\G(\omega)\equiv\G(00,\omega)$.  Combined with the Dyson
equation $\G^r=\G^r_0+\G^r_0\Sigma^r\G^r$ at $x=x'=0$, we
obtain
\begin{eqnarray}
\theta(\omega)\textrm{Re}\Big[1+\G^{r}(\omega)\Sigma^{r}(\omega)\Big]
=\theta(\omega)\frac{A(\omega)}{A_0(\omega)}\nonumber
\end{eqnarray}
in terms of the local spectral function
$A(\omega)=i[\G^r(\omega)-\G^a(\omega)]$. After multiplication
with $-e<0$, we finally arrive at the electric current in
(\ref{current}).  Note that the current formula (\ref{current})
can be seen as a continuous-space version of the Meir-Wingreen
formula\cite{Meir-Wingreen} for transport through an Anderson
dot, see \cite{lundethesis} for a detailed discussion.


\section*{References}

\begin{thebibliography}{99}

\bibitem{wharam}
D.A. Wharam, T.J. Thornton, R. Newbury, M. Pepper, H. Ahmed, J.E.F. Frost, D.G. Hasko, D.C. Peacock, D.A. Ritchie, and G.A.C. Jones, J. Phys. C {\bf 21}, L209 (1988); B.J. van Wees, H. van Houten, C.W.J. Beenakker, J.G. Williamson, L.P. Kouwenhoven, D. van der Marel, and C.T. Foxon, Phys. Rev. Lett. {\bf 60}, 848 (1988).

\bibitem{thomas}
K.J. Thomas, J.T. Nicholls, M.Y. Simmons, M. Pepper, D.R. Mace,
and D.A. Ritchie, Phys. Rev. Lett. {\bf 77}, 135 (1996).

\bibitem{kristensen} A. Kristensen, H. Bruus, A.E. Hansen, J.B. Jensen, P.E. Lindelof, C.J. Marckmann, J. Nyg\aa rd, C.B. S{\o}rensen, F. Beuscher, A. Forchel, and M. Michel, Phys. Rev. B {\bf 62}, 10950 (2000).

\bibitem{cronenwett}
S.M. Cronenwett, H.J. Lynch, D. Goldhaber-Gordon, L.P. Kouwenhoven, C.M. Marcus, K. Hirose, N.S. Wingreen, and V. Umansky, Phys. Rev. Lett.  {\bf 88}, 226805
(2002).

\bibitem{appleyard}
N.J. Appleyard, J.T. Nicholls, M. Pepper, W.R. Tribe, M.Y. Simmons,
and D.A. Ritchie, Phys. Rev. B {\bf 62}, R16275 (2000).

\bibitem{thermo2}
J.T. Nicholls and O. Chiatti,
J. Phys.: Condens. Matter {\bf 20}, 164210 (2008).


\bibitem{roche}
P. Roche, J. S{\'e}gala, D.C. Glattli, J.T. Nicholls, M. Pepper, A.C. Graham, K.J. Thomas, M.Y. Simmons, and D.A. Ritchie,  Phys. Rev. Lett. {\bf 93}, 116602 (2004); L. DiCarlo, Y. Zhang, D.T. McClure, D.J. Reilly, C.M. Marcus, L.N. Pfeiffer, and K.W. West, Phys Rev. Lett. {\bf 97}, 036810 (2006).

\bibitem{sfigakis}
F. Sfigakis, C.J.B. Ford, M. Pepper, M. Kataoka, D.A. Ritchie, and M.Y. Simmons, Phys. Rev. Lett. {\bf 100}, 026807 (2008).

\bibitem{bruus}
H. Bruus, V.V. Cheianov, and K. Flensberg, Physica E {\bf 10}, 97
(2001); D.J. Reilly, Phys. Rev. B {\bf 72}, 033309  (2005).

\bibitem{richter}
A. Lassl, P. Schlagheck, and K. Richter, Phys. Rev. B {\bf 75},
045346 (2007).

\bibitem{zou}
S. Ihnatsenka and I.V. Zozoulenko, Phys. Rev. B {\bf 76}, 045338 (2007).

\bibitem{bulka}
B.R. Bulka, T. Kostyrko, M. Tolea, and I.V. Dinu, J. Phys.: Condens. Matter
{\bf 19}, 255211 (2007).

\bibitem{meir}
Y. Meir, K. Hirose, and N.S. Wingreen, Phys. Rev. Lett. {\bf 89},
196802 (2002); T. Rejec and Y. Meir, Nature {\bf 442}, 900 (2006).

\bibitem{cornagliabalseiro}
P.S. Cornaglia and C.A. Balseiro, Europhys. Lett. \textbf{67}, 634
(2004); P.S. Cornaglia, C.A. Balseiro, and M. Avignon, Phys.
Rev. B {\bf 71}, 024432 (2005).

\bibitem{berggren}
C.K. Wang and K.-F. Berggren, Phys. Rev. B {\bf 54}, R14257
(1996); A.A. Starikov, I.I. Yakimenko, and K.-F. Berggren, Phys.
Rev. B {\bf 67}, 235319 (2003); K.-F. Berggren and I.I. Yakimenko,
J. Phys.: Condens. Matter {\bf 20}, 164203 (2008).

\bibitem{seeligmatveev} G. Seelig and K.A. Matveev, Phys. Rev. Lett.
{\bf 90}, 176804 (2003)

\bibitem{matveev} K.A. Matveev, Phys. Rev. Lett. {\bf 92}, 106801 (2004);
M. Kindermann and P.W. Brouwer, Phys. Rev. B {\bf 74}, 125309 (2006).

\bibitem{meidanoreg}
D. Meidan and Y. Oreg, Phys. Rev. B \textbf{72}, 121312(R) (2005).

\bibitem{schmeltzer}
D. Schmeltzer, A. Saxena, A.R. Bishop, and D.L. Smith, Phys. Rev. B
\textbf{71}, 045429 (2005).

\bibitem{syljuaasen}
O.F. Sylju\aa sen, Phys. Rev. Lett. \textbf{98}, 166401  (2007).

\bibitem{sushkov}
C. Sloggett, A.I. Milstein, and O.P. Sushkov,  Eur. Phys. J. B \textbf{61}, 427 (2008).

\bibitem{matv2}
J. Rech and K.A. Matveev, Phys. Rev. Lett. \textbf{100}, 066407 (2008).

\bibitem{preprint}
A.M. Lunde, A. De Martino, R. Egger, and K. Flensberg,
cond-mat/0707.1989.

\bibitem{newfoot} Note that even in the long-wire limit, an
    inhomogeneous interaction can cause a resistivity change,
    see \cite{matv2}.

\bibitem{glazman88}
L.I. Glazman and R.I. Shekhter, Zh. Eksp. Teor. Fiz. {\bf 94}, 292
(1988) [Sov.~Phys.  JETP {\bf 67}, 163 (1988)].

\bibitem{buttiker86}
M. B\"uttiker, Phys. Rev. Lett. {\bf 57}, 1761 (1986).

\bibitem{footnote:new} The new feature for QPCs comes from the
    momentum-nonconserving interactions, resulting in the
    distinct low-to-intermediate temperature dependence $G(T)$.
    The perturbative $T^2$ correction is not present in the
    published long wire results\cite{matveev}, but is expected
    because of the momentum-nonconserving scattering taking
    place at the ends of the wire.

\bibitem{buttiker}
M. B\"uttiker, Phys. Rev. B {\bf 41}, 7906 (1990).

\bibitem{glazman88-adiabatisk-QPC}
L.I. Glazman, G. B. Lesovik, D.E. Khmelnitskii and R.I. Shekhter, JETP. Lett {\bf 48}, 238 (1988).

\bibitem{gry}
I.S. Gradsteyn and I.M. Ryzhik, {\sl Table of Integrals, Series, and
Products} (Academic Press, Inc., New York, 1980).

\bibitem{Taboryski1995}
R. Taboryski, A. Kristensen, C. B. S{\o}rensen, and P. E. Lindelof, Phys. Rev. B
\textbf{51},  2282 (1995).

\bibitem{Jauho-book}
H. Haug and A.-P. Jauho, \emph{Quantum Kinetics in Transport and Optics of Semiconductors}, 1st ed. (Springer, New York, 1996).

\bibitem{Rammer-smith1986}
J. Rammer and H. Smith, Rev. Mod. Phys. \textbf{58}, 323 (1986).


\bibitem{lundeprl}
A.M. Lunde, K. Flensberg, and L.I. Glazman, Phys. Rev. Lett.
\textbf{97}, 256802 (2006).

\bibitem{lundeprb2007}
A.M. Lunde, K. Flensberg, and L.I. Glazman, Phys. Rev. B
\textbf{75},  245418 (2007).

\bibitem{saleur}
P. Fendley and H. Saleur, Phys. Rev. B {\bf 54}, 10845 (1996);
V.V. Ponomarenko and N. Nagaosa, Phys. Rev. B {\bf 60}, 16865 (1999).

\bibitem{thermodef}
H. van Houten, L.W. Molenkamp, C.W.J. Beenakker, and
C.T. Foxon, Semicond. Sci. Technol. {\bf 7}, B215 (1992).

\bibitem{gogolin}
A.O. Gogolin, A.A. Nersesyan, and A.M. Tsvelik, {\sl Bosonization and Strongly Correlated Systems}
(Cambridge University Press, 1998).

\bibitem{footnote:lambda}
This follows by matching the perturbative result (\ref{currpert})
to the corresponding perturbative result for the point-like interaction model,
$G^{(2)}/G_0=- (2\pi/3) [\pi\lambda T/T_F]^2$.

\bibitem{hershfield}
S. Hershfield, J.H. Davies, and J.W. Wilkins, Phys. Rev. B {\bf 46}, 7046 (1992).

\bibitem{lundethesis}
A.M. Lunde, PhD Thesis (Copenhagen, 2007).

\bibitem{glazman}
For the noninteracting case, this is discussed by L.I. Glazman and M. Jonson, Phys. Rev. B {\bf 44}, 3810 (1991).

\bibitem{Meir-Wingreen}
Y. Meir and N.S. Wingreen, Phys. Rev. Lett. {\bf 68}, 2512 (1992).


\end{thebibliography}
\end{document}